\documentclass[12pt]{iopart}
\usepackage [utf8]{inputenc}
\usepackage{iopams}
\expandafter\let\csname equation*\endcsname\relax

\expandafter\let\csname endequation*\endcsname\relax

\usepackage{bm,bbm}
\usepackage{amsmath}
\usepackage{ams fonts}
\usepackage{amstext,amsthm}
\usepackage{amssymb}   
\usepackage{graphicx}
\usepackage{graphics}
\usepackage{verbatim}
\usepackage{cite}
\usepackage{caption}
\usepackage[labelformat=simple]{subcaption}
\usepackage{xcolor,etoolbox}
\makeatletter 
\pretocmd\@bibitem{\color{black}\csname keycolor#1\endcsname}{}{\fail}

\begin{document}

\hyphenation{ge-ne-ra-li-zed}
\def\hf{\hat{f}}
\def\Ord{\mathcal{O}}
\newcommand{\barr}{\begin{eqnarray}}
\newcommand{\earr}{\end{eqnarray}}
\newcommand{\beq}{\begin{equation}}
\newcommand{\eeq}{\end{equation}}
\newcommand{\be}{\begin{equation}}
\newcommand{\ee}{\end{equation}}
\newcommand{\ra}{\right\rangle}
\newcommand{\de}{\mathrm{d}}
\newcommand{\la}{\left\langle}
\newcommand{\correc}{ \textcolor{red} }
\newtheorem{theorem}{Theorem}

\newcommand{\gv}[1]{\ensuremath{\mbox{\boldmath$ #1 $}}} 
\newcommand{\uv}[1]{\ensuremath{\bm{\hat{#1}}}} 
\newcommand{\abs}[1]{\left| #1 \right|} 
\let\underdot=\d 
\renewcommand{\d}[2]{\frac{d #1}{d #2}} 
\newcommand{\dd}[2]{\frac{d^2 #1}{d #2^2}} 
\newcommand{\pd}[2]{\frac{\partial #1}{\partial #2}} 
\newcommand{\pdd}[2]{\frac{\partial^2 #1}{\partial #2^2}} 
\newcommand{\pdc}[3]{\left( \frac{\partial #1}{\partial #2}
 \right)_{#3}} 
\newcommand{\ket}[1]{\left| #1 \right>} 
\newcommand{\bra}[1]{\left< #1 \right|} 
\newcommand{\braket}[2]{\left< #1 \vphantom{#2} \right|
 \left. #2 \vphantom{#1} \right>} 
\newcommand{\matrixel}[3]{\left< #1 \vphantom{#2#3} \right|
 #2 \left| #3 \vphantom{#1#2} \right>} 
\newcommand{\grad}[1]{\gv{\nabla} #1} 
\let\divsymb=\div 
\renewcommand{\div}[1]{\gv{\nabla} \cdot #1} 
\newcommand{\curl}[1]{\gv{\nabla} \times #1} 
\let\baraccent=\= 
\renewcommand{\=}[1]{\stackrel{#1}{=}} 

\newcommand{\numberset}{\mathbb}
\newcommand{\N}{\numberset{N}}
\newcommand{\Z}{\numberset{Z}}
\newcommand{\R}{\numberset{R}}
\newcommand{\C}{\numberset{C}}

\newcommand{\Res}{\mathrm{Res}}
\newcommand{\Var}{\mathrm{Var}}
\newcommand{\Cov}{\mathrm{Cov}}
\newcommand{\avg}[1]{\left< #1 \right>} 
\newcommand{\T}{\mathcal{T}}
\newcommand{\rs}{\rho^{\star}}
\def\im{{\rm i}}
\newcommand{\kk}{\kappa}
\newcommand{\EE}{\mathcal{E}}
\newcommand{\Cc}{\mathcal{C}}

\def\lm{\lambda_-}
\def\lp{\lambda_+}
\def\lpm{\lambda_\pm}

\def\Wt{\tau_\mathrm{W}}
\def\Ht{\tau_\mathrm{H}}
\def\Sm{\mathcal{S}}

\def\Ht{\tau_\mathrm{H}}
\def\Nc{N}
\def\Sm{\mathcal{S}}
\def\rho{\varrho}

\providecommand{\keywords}[1]{\textbf{\textit{Keywords: }} #1}

\title[A memory-based method to select the number of relevant components  in PCA]{A memory-based method to select the number of relevant components in Principal Component Analysis}

\author{Anshul Verma$^1$, Pierpaolo Vivo$^1$ and Tiziana Di Matteo$^{1,2,3}$}

\address{$^1$ Department of Mathematics, King's College London, Strand, London, WC2R 2LS, United Kingdom\\
$^2$ Department of Computer Science, University College London, Gower Street, London, WC1E 6BT, United Kingdom\\
$^3$ Complexity Science Hub Vienna, Josefst\"{a}dter Strasse 39, A 1080 Vienna, Austria}
\ead{anshul.verma@kcl.ac.uk, pierpaolo.vivo@kcl.ac.uk, tiziana.di\_matteo@kcl.ac.uk}

\begin{abstract}
We propose a new data-driven method to select the optimal number of relevant components in Principal Component Analysis (PCA). This new method applies to correlation matrices whose time autocorrelation function decays more slowly than an exponential, giving rise to long memory effects. In comparison with other available methods present in the literature, our procedure does not rely on subjective evaluations and is computationally inexpensive. The underlying basic idea is to use a suitable factor model to analyse the residual memory after sequentially removing more and more components, and stopping the process when the maximum amount of memory has been accounted for by the retained components. We validate our methodology on both synthetic and real financial data, and find in all cases a clear and computationally superior answer entirely compatible with available heuristic criteria, such as cumulative variance and cross-validation.
\end{abstract}


\maketitle

\tableofcontents

\section{Introduction}
With the arrival of sophisticated new technologies and the advent of the Big Data era, the amount of digital information that can be produced, processed and stored has increased at an unprecedented pace in recent years. The need of sophisticated post-processing tools -- able to identify and discern the essential driving features of a given high-dimensional system -- has thus become of paramount importance. Principal Component Analysis (PCA), which aims to reduce the dimensionality of the correlation matrix between data \cite{Jolliffe2002,Jackson1993}, is continuing to prove a highly valuable method in this respect. PCA has been shown to have applications spanning from neuroscience to finance. In image processing, for instance, this technique has proven useful to identify key mixtures of colours of an image for use in compression \cite{Sonka2014}. In molecular dynamics, the increasing computational power available to researchers makes it possible to simulate more complex systems, with PCA helping to detect important chemical drivers \cite{Stein2006}. The brain's neurons produce different responses to a variety of stimuli, hence PCA can be used in neuroscience to find common binding features that determine such responses \cite{Pang2016}. In finance, the amount of digital storage and the length of available historical time series have dramatically increased. It has therefore become possible to probe the multivariate structure of changes in prices, but with the large universe of stocks that usually make up markets, PCA has become a valuable technique in identifying essential factors governing price evolution \cite{Alexander2002,Bun2017,Darbyshire2016}.

Within the class of \emph{dimensionality reduction} methods, whose goal is to produce a faithful but smaller representation of the original correlation matrix \cite{Maaten2009}, PCA plays a very important role. Other known methods include information filtering techniques \cite{Aste2005,Tumminello2005,Matteo2010,Aste2012,Barfuss2016,Verma2017}, autoencoders \cite{Hinton1994,Regier2016} and Independent Component Analysis (ICA) \cite{Comon1994,Hyvarinen2004}. PCA accomplishes this task using a subset of the orthogonal basis of the correlation matrix of the system. Successive \emph{principal components} -- namely the eigenvectors corresponding to the largest eigenvalues -- provide the orthogonal directions along which data are maximally spread out. Since the dimension of empirical correlation matrices can be as large as $\sim 10^{2}-10^3$, a highly important parameter is the number $m^\star$ of principal components one should retain, which should strike the optimal balance between providing a faithful representation of the original data and avoiding the inclusion of irrelevant details. 

Unfortunately, there is no natural prescription on how to select the optimal value $m^\star$, and many heuristic procedures and so-called \emph{stopping criteria} have been proposed in the literature \cite{Jackson1993,Jolliffe2002}. The most popular methods -- about which more details are given in Section \ref{Comparison}) --  are i) scree plots \cite{Cattell1966}, ii) cumulative explained variance \cite{Sugiyama1976,Huang1992}, iii) distribution-based methods \cite{Bartlett1950,Ali1985}, and iv) cross-validation \cite{Wold1978,Eastment1982}. However, they all suffer from different, but serious drawbacks: i) and ii) are essentially rules of thumb with little data-driven justification, iii) do not allow the user to control the overall significance level of the final result and are thus impractical for large data sets, and finally iv), whilst being more objective and relying on fewer assumptions, is often computationally cumbersome \cite{Jolliffe2002}. Efforts to improve each subclass -- for instance the more ``subjective" methods \cite{Cattell1966,Sugiyama1976,Huang1992} -- have been undertaken, but they usually resulted in adding more assumptions or were anyway unable to fully solve the issues \cite{Jolliffe2002}. 

Unlike most other methods available in the literature, in this paper we propose to take advantage of long memory effects that are present in many empirical time series \cite{Beran2017} to select the optimal number $m^\star$ of principal components to retain in PCA. We shall leverage on the natural factor model implied by PCA (see Section \ref{Regression} below) to assess the statistical contribution of each principal component to the overall ``total memory" of the time series, using a recently introduced proxy for memory strength \cite{Verma2017}. We test the validity of our proposal on synthetic data, namely two fractional Gaussian noise processes with different Hurst exponents (see Section \ref{Synthetic}), and also on an empirical dataset whose details are reported in \ref{DataCleaning}. Comparing our memory-based method with other heuristic criteria in the literature, we find that our procedure does not include any subjective evaluation, makes a very minimal and justifiable set of initial assumptions, and is computationally far less intensive than cross-validation. 

Our methodology is generally applicable to any (however large) correlation matrix of a long-memory dataset. A typical example is provided by financial time-series, which are well-known to display long-memory effects \cite{Cont2001}. The volatility of such time-series indeed constitutes an important input for risk estimation and dynamical models of price changes \cite{Hull1987,Hull2006,Bouchaud2009b}. However, the multivariate extensions of common volatility models, such as multivariate Generalised Autogressive Conditional Heteroskedastic (GARCH) \cite{Bauwens2006}, stochastic covariance \cite{Clark1973} and realised covariance \cite{Andersen2003}, suffer from the curse of dimensionality, hindering their application in practice. A popular solution to this issue is to first apply PCA to the correlation matrix between volatilities, and then use the reduced form of the correlation matrix to fit a univariate volatility model for each component,  as in \cite{Alexander2002}. In climate studies, PCA has been used to create `climate indices' to identify patterns in climate data from a wide range of measurements including precipitations and temperature \cite{Von2001}. Here, factors such as the surface temperature are known to exhibit long range memory \cite{Franzke2012}. In neuroscience, PCA can be used to discover amongst the vast number of possible neurons those which correspond to particular responses, for example how an insect brain responds to different odorants \cite{Pang2016}. In this case as well, long memory effects are well-known to play an important role \cite{Linkenkaer2001}. Our framework is therefore highly suited to a wide array of problems. 

The paper is organised as follows: in Section \ref{PCAIntroduction}, we introduce and define the PCA procedure and how one selects the most relevant number of principal components. Section \ref{FinancialData} describes the relevant quantities and results that are specific to financial data. We detail our proposed method to select the principal components based on memory in Section \ref{Method}, testing the method on synthetic and empirical data in Section \ref{ApplyingMethod}. We explore the advantages that our method offers over existing approaches in literature in Section \ref{Comparison}, before finally drawing some conclusions in Section \ref{Conclusion}. The appendices are devoted to the description of the empirical dataset and technical details.

\section{PCA and the optimal number of principal components to retain} \label{PCAIntroduction}
In this Section, we give a brief introduction to PCA to make the paper self-contained. Call $\bm{X}$ the data matrix, which contains $N$ columns -- standardised to have zero mean and unit variance -- of individual defining features, and $T$ rows recording particular realisations in time of such features. PCA searches for the orthogonal linear basis with unit length $\bm{w}_{\{i=1,...,N\}}$ that transforms the system to one where the highest variance is captured by the first component, the second highest by the second component and so on \cite{Jolliffe2002}. The first component is therefore given by
\begin{equation}
\bm{w}_{1}=\arg \max_{||\bm{w}||=1}\left\{||\bm{X}\bm{w}||^{2}\right\}=\arg \max_{||\bm{w}||=1}\left\{\bm{w}^{\dagger}\bm{E} \bm{w}\right\} \ ,
\end{equation}
where $\dagger$ represents the transpose, and $\bm{E}$ is the sample correlation matrix of $\bm{X}$, defined as
\begin{equation}
E_{ij}=\frac{1}{T}\sum_{t=1}^{T}X_{ti}X_{tj} \ . \label{EEqn}
\end{equation}
The search for $\bm w_1$ can be formulated as a constrained optimisation problem, i.e. we must maximise
\begin{equation}
\bm{w}^{\dagger}\bm{E} \bm{w}-\lambda(\bm{w}^{\dagger}\bm{w}-1) \ , \label{Lagrange}
\end{equation}
where $\lambda$ is the Lagrange multiplier enforcing normalisation of the eigenvectors. Differentiating Eq. \eqref{Lagrange} w.r.t. to $\bm{w}$ we get
\begin{equation}
\bm{E}\bm{w}-\lambda\bm{w}=0 \ .
\end{equation} 
This means that the Lagrange multiplier must be an eigenvalue of $\bm{E}$. Also note that the variance of data along the direction $\bm w$ is given by
\begin{equation}
\bm{w}^{\dagger}\bm{E}\bm{w}=\lambda\bm{w}^{\dagger}\bm{w}=\lambda \ ,
\end{equation}
and hence the largest variance is realised by the top eigenvalue. It follows that the first principal component -- i.e. the direction along which the data are maximally spread out -- is nothing but the top eigenvector $\bm{w}_{1}$ corresponding to the top eigenvalue $\lambda_1$. A similar argument holds for the subsequent principal components. 

The aim of PCA is to reduce $\bm{E}$ to a $m \times m$ matrix, where $m\ll N$ is the number of principal components that we choose to retain. Is there an optimal value $m^\star$ that one should select? Clearly, this is an important question that must be addressed, since it determines the ``best" size of the reduced correlation matrix that is just enough to describe the main features of the data without including irrelevant details. In this paper, we address this question and we provide a new method to select the optimal value $m^\star$ of the number of principal components that we should retain for long-memory data.

\section{Financial Data} \label{FinancialData}
\subsection{Data Structure} \label{DataStructure}
In this Section, we describe the general structure of the data matrix that we use in the context of financial data. We consider a system of $N$ stocks and $T$ records of their daily closing prices. We calculate the time series of log-returns for a given stock $i$, $r_{i}(t)$, defined as: 
\begin{equation}
r_{i}(t)=\ln p_{i}(t+1) - \ln p_{i}(t) \ ,
\end{equation}
where $p_{i}(t)$ is the price of stock $i$ at time $t$. After standardising $r_{i}(t)$ so that it has zero mean and unit variance, we define the proxy we shall use for the volatility, i.e. the variability in asset returns (either increasing or decreasing), as $\ln |r_{i}(t)|$ \cite{Taylor1994}. Most stochastic volatility models -- where the volatility is assumed to be random and not constant -- assume that the return for the stock $i$ evolve according to \cite{Breidt1998}
\begin{equation}
r_{i}(t)=\delta(t)\exp^{\omega_{i}(t)} \ , \label{StochVolModel}
\end{equation}
where $\delta(t)$ is a white noise with finite variance and $\omega_{i}(t)$ are the \emph{log volatility} terms. The exponential term encodes the structure of the volatility and how it contributes to the overall size of the return. We note that for our purposes, we are able to set the white noise term to be the same for all stocks since it contains no memory by definition \cite{Box2015} (we have checked that changing this assumption to include a stock dependent white noise term does not change our results). Taking the absolute value of Eq. \ref{StochVolModel} and the log of both sides, Eq. \ref{StochVolModel} becomes
\begin{equation} \label{LogVolTerms}
\ln |r_{i}(t)|=\ln |\delta(t)|+\omega_{i}(t) \ .
\end{equation}
We see that working with $\ln |r_{i}(t)|$ has the added benefit of making $\omega_{i}(t)$ -- the proxy for volatility -- additive, which in turn makes the volatility more suitable for factor models. Since $\delta(t)$ is a random scale factor that is applied to all stocks, we can set it to $1$, so that $\omega_{i}(t)=\ln |r_{i}(t)|$. We also standardise $\omega_{i}(t)$ to a mean of $0$ and standard deviation $1$ as performed in \cite{Singh2016}. Finally, call $\mathbf{X}$ the data matrix, which contains $N$ columns for each individual defining stock, and $T$ rows recording particular realisations in time of such stocks so that the $i,t$ entry of $\mathbf{X}$ is $X_{it}=\omega_{i}(t)$. 

\subsection{Market Mode and Mar\v{c}enko-Pastur} \label{MarchenkoPastur}

For the case of log volatilities in finance \cite{Schafer2010,Bun2017} (see further details in \ref{PortfolioOpt}), it has been known for some time that the smallest eigenvalues of the empirical correlation matrix $\bm E$ may be heavily contaminated by noise due to the finiteness of the data samples. In our search for the most relevant $m^\star$ components, it is therefore important to confine ourselves to the sector of the spectrum that is less affected by noise at the outset.

To facilitate this identification, we will resort to a null distribution of eigenvalues, which are produced from a Gaussian white noise process. This is given by the celebrated Mar\v{c}enko-Pastur (MP) distribution \cite{Marchenko1967,Bun2017,Livan2018} 
\begin{equation}
p(\lambda)=\frac{1}{2\pi q\sigma^{2}}\frac{\sqrt{(\lambda_{+}-\lambda)(\lambda-\lambda_{-})}}{\lambda} \label{MPDist} \ ,
\end{equation}
where $p(\lambda)$ is the probability density of eigenvalues having support in $\lambda_{-}< \lambda < \lambda_{+}$. The edge points $\lambda_{\pm}=\sigma\left(1\pm \sqrt{q}\right)^{2}$, $q=T/N$ and $\sigma$ is the standard deviation over all stocks. By comparing the empirical eigenvalue distribution of $\bm E$ to the MP law \eqref{MPDist}, we can therefore see how many eigenvalues, and thus principal components, are likely to be corrupted by noise and should therefore be discarded from the very beginning. More recently, this procedure has received some criticisms \cite{Guhr2003,Livan2011,Wilinski2018}: it has been argued that eigenvalues carrying genuine information about weakly correlated clusters of stocks could still be buried under the MP sea, and more refined filtering strategies may be needed to bring such correlations to the surface. Other generalisations of the MP law for non-normally distributed random data and applications to financial data can be found for instance in \cite{Biroli2007} and \cite{Abul2009}.  

\begin{figure}
\begin{subfigure}{0.475\textwidth}
\centering
\includegraphics[width=\textwidth]{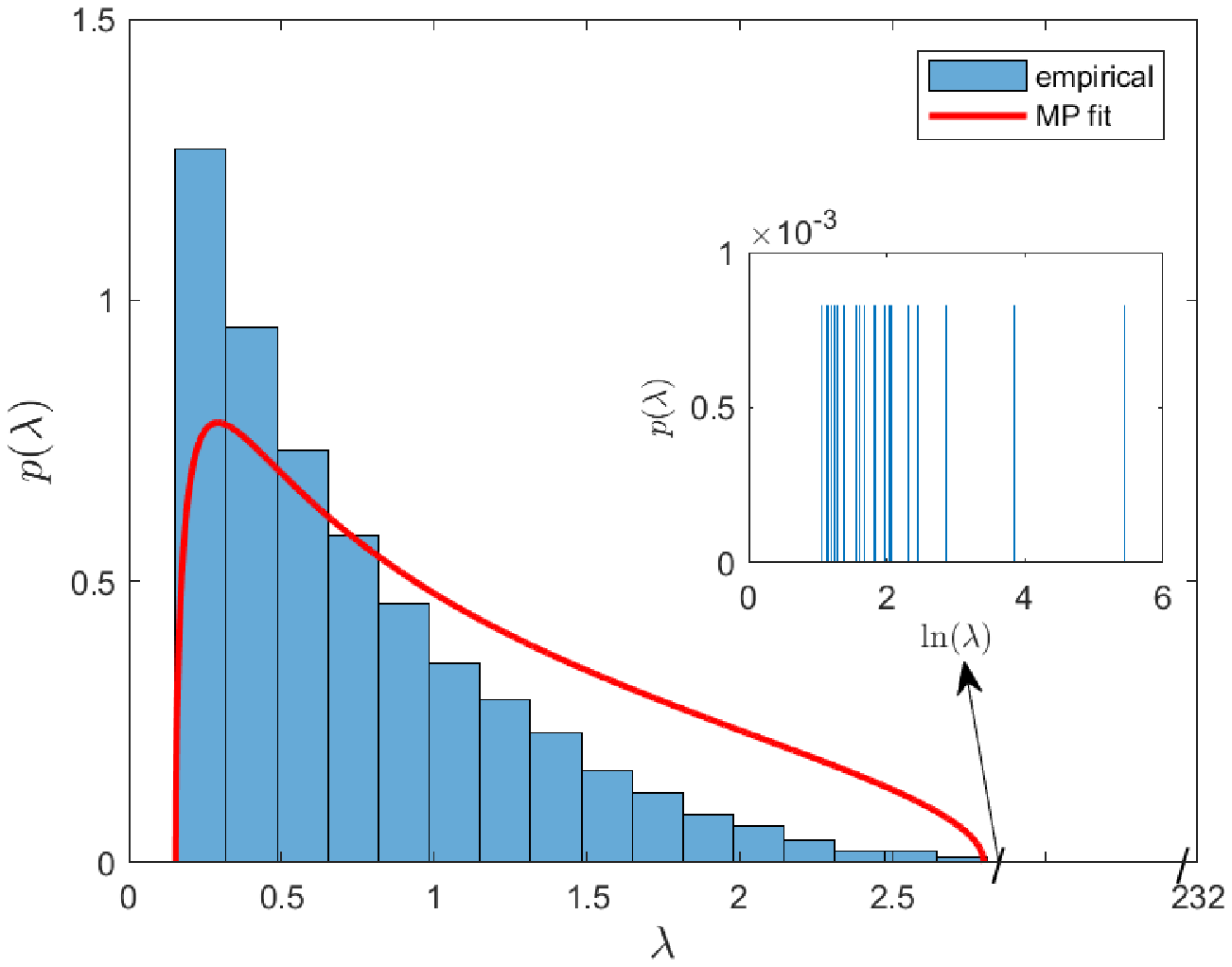}
\caption{\label{fig:EigenvalueSpec}\qquad $\bm{E}$}
\end{subfigure}
\begin{subfigure}{0.475\textwidth}
\centering
\includegraphics[width=\textwidth]{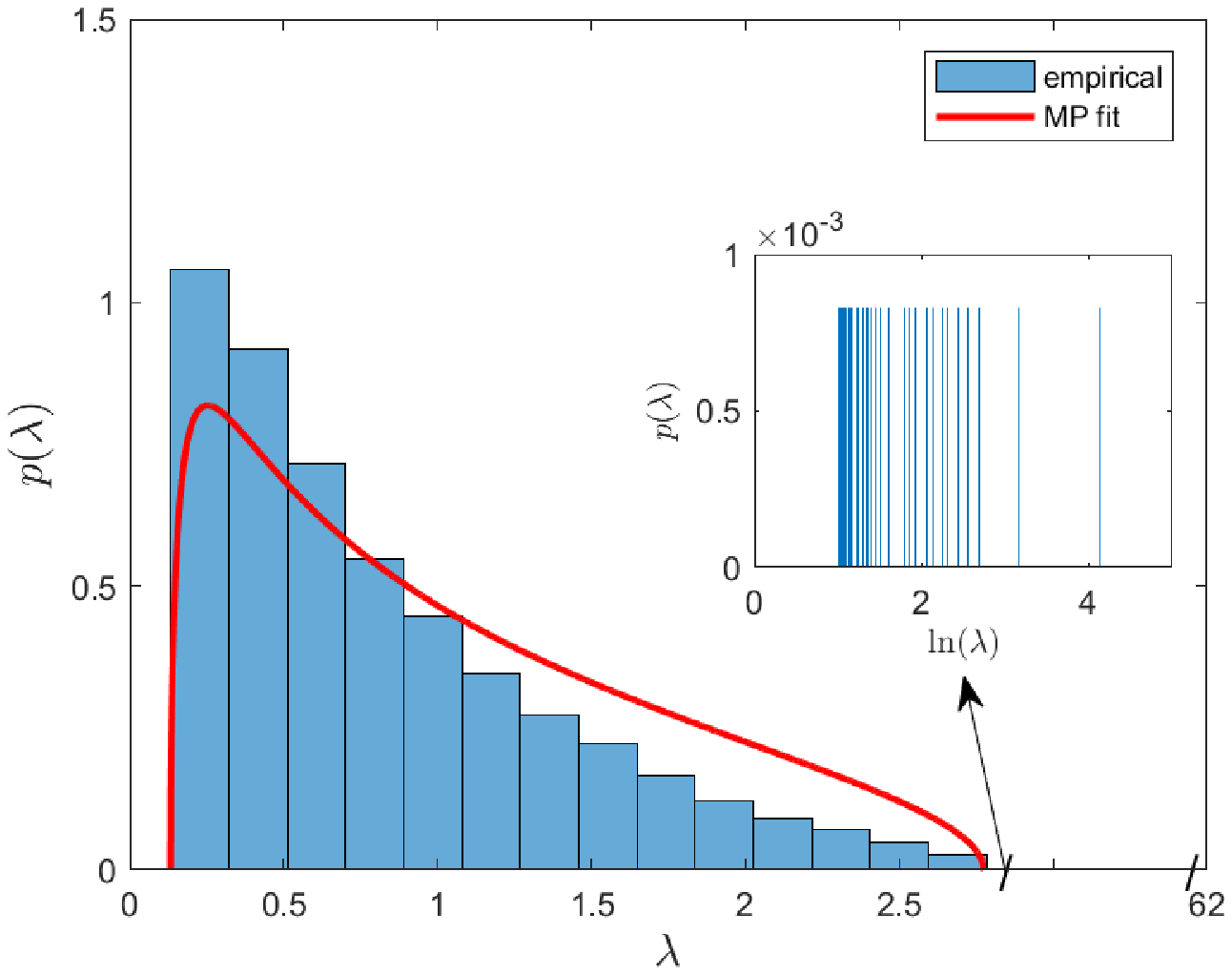}
\caption{\label{fig:EigenvalueSpec_detrend}\qquad $\bm{G}$}
\end{subfigure}
\caption{(a) Histogram of the eigenvalue distribution of $\bm{E}$ constructed from the empirical dataset (see \ref{DataCleaning}), compared to the best fit Mar\v{c}enko-Pastur distribution in red. The $\lambda$ axis has been split by the forward-slashes to only show the bulk eigenvalues below $\lambda_{+}=2.80$. The inset shows the $22$ isolated eigenvalues for $\lambda>\lambda_{+}$ in semilog scale. The Mar\v{c}enko-Pastur distribution is fitted with parameters $q=0.38\pm 0.02$ and $\sigma =1.03\pm 0.01$. (b) Same histogram, but applied to the correlation matrix $\bm{G}$ (see Section \ref{DetrendMarketMode}), where the market mode has been de-trended. Here $\lambda_{+}=2.77$, $q=0.41\pm 0.02$, $\sigma=1.01\pm 0.01$, with $35$ eigenvalues above $\lambda_{+}$. }
\label{fig:MarchenkoPastur}
\end{figure}

\begin{figure}
\centering
\includegraphics[width=0.75\textwidth]{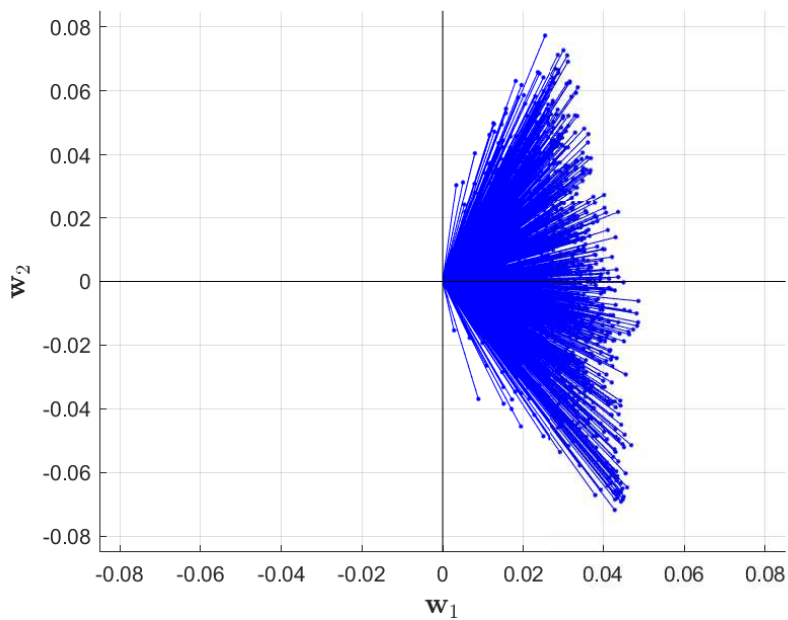}
\caption{Each point $i$ in this graph has coordinates $(w_{i1},w_{i2})$, where $w_{i1}$ is the $i$-th entry of the top eigenvector $\bm{w}_{1}$, and $w_{i2}$ is the $i$-th entry of the second eigenvector $\bm{w}_{2}$ of the correlation matrix $\bm{E}$ (defined in Eq. \eqref{EEqn}) for the dataset in \ref{DataCleaning}. The length of the corresponding 2D coordinate vector from the origin is given by each blue line. The plot shows that all values of $w_{i1}$ are in fact positive.}
\label{fig:BiPlot}
\end{figure}

In practical terms, we first create the empirical correlation matrix $\bm E=(1/T)\bm X^\dagger\bm X$ from the matrix $\bm X$ (constructed from either synthetic or empirical data), and then we fit the MP law to the empirical distribution of its eigenvalues. This is done by considering $q$ and $\sigma$ in Eq. \eqref{MPDist} as free parameters to take into account finite sample biases \cite{Livan2011}. In Fig. \ref{fig:EigenvalueSpec}, we plot the histogram of bulk eigenvalues for the empirical dataset described in \ref{DataCleaning}, and in the inset a number of outliers $\lambda > \lambda_{+}$ in semilog scale. It is indeed well-known that some of the eigenvalues of $\bm E$ extend well beyond the upper edge of the MP law, and that the largest eigenvalue lies even further away (see Fig. \ref{fig:EigenvalueSpec}). This means that the first principal component accounts for a large proportion of the variability of data, and is in fact a well-known effect of the \emph{market mode} \cite{Laloux1999,Plerou2002,Singh2016}. We plot the entries of the right eigenvector $\bm{w}_{1}$ of $\mathbf{E}$ (corresponding to the market mode) and $\bm{w}_{2}$ in Fig. \ref{fig:BiPlot}, with the blue lines giving the length from the origin of the corresponding 2D vector. We see from Fig. \ref{fig:BiPlot} that the entries for $\bm{w}_{1}$ are all positive, which confirms that indeed the first eigenvector affects all stocks.

\section{Long Memory} \label{VolatilityClustering}
We now consider the `long memory' features of a time series, specialising the discussion to the log volatility in a financial context. 

The autocorrelation function (ACF), $\kappa(L)$, of any time series $x(t)$ is defined as
\begin{equation}
\kappa(L)=\mathrm{corr}(x(t+L) ,x(t)) =\frac{\langle\left[x(t+L) x(t)\right]\rangle}{\sigma^{2}} \ , \label{ACF}
\end{equation}
where $\langle...\rangle$ denotes the time expectation over $x(t)$, adjusted to have zero mean. $L$ is the lag and $\sigma^{2}$ is the variance of the process $x(t)$. If $\kappa(L)$ decays faster than or as fast as an exponential with $L$, then the time series is said to have short memory \cite{Beran2017}. However, in many real world systems ranging from outflows in hydrology to tree ring measurements \cite{Beran2017}, $\kappa(L)$ has been found to decay much more slowly than an exponential,  giving rise to an important effect known as long memory \cite{Beran2017}. This means that the process at time $t$ remains heavily influenced by what happened in a rather distant past. In particular for financial data (where $x(t)=|\ln r(t)|$), it is an accepted stylised fact (called \emph{volatility clustering}) that large changes in volatilities are usually followed by other large changes in volatilities, or that the volatilities retain a long memory of previous values \cite{Mandelbrot1997}. $\kappa(L)$ has also been empirically found to follow a power law decay
\begin{equation}
\kappa(L)\sim L^{-\beta^{\text{vol}}} \ , \label{PowerLawACF}
\end{equation}
where $\beta^{\text{vol}}$ describes the strength of the memory effect -- a lower value indicates that a longer memory of past values is retained.
However, as shown in \cite{Verma2017}, to better distinguish between short and long memory it is convenient to consider the non-parametric integrated proxy $\eta$, defined as
\begin{equation}
\eta=\int_{L=1}^{L_{cut}}\kappa(L)\de L \ ,\label{IntProxy}
\end{equation}
where $L_{cut}$ is the standard Bartlett Cut at the 5\% level \cite{Box2015}. The proxy $\eta$ is less affected by the noise-dressing of $\kappa(L)$ than $\beta^{\text{vol}}$ \cite{Verma2017}, and the larger the value of $\eta$ the greater the degree of the memory effect. This observable will constitute an essential ingredient of our method.

\section{Methods} \label{Method}

In this Section, we describe in detail our procedure.

\subsection{De-trending the market mode} \label{DetrendMarketMode}
The first step of our method consists in removing the influence of the market mode, the global factor affecting the data, as we hinted at in Section \ref{MarchenkoPastur}. To do this, we impose that the standardised log volatility $\omega_{i}(t)=\ln |r_{i}(t)|$ in Eq. \eqref{LogVolTerms} follows a factor model (using the Capital Asset Pricing Model, CAPM \cite{Sharpe1964,Merton1973}) with the \emph{market mode}
\begin{equation}
I_{0}(t)=\sum_{i=1}^{N}w_{i1}\ln |r_{i}(t)|  \label{MarketModeWeighted}
\end{equation}
as a factor. This quantity -- essentially a weighted average of $\ln |r_{i}(t)|$ with weights $\bm{w}_{1}$, the top eigenvector's components -- represents the effect of the market as a whole on all stocks i.e. the common direction taken by all stocks at once.

Hence we define
\begin{equation}
\omega_{i}(t)=\beta_{i0}I_{0}(t)+\alpha_{i0}+c_{i}(t) \ .\label{MarketEqn}
\end{equation}
Here, $\beta_{i0}$ is the responsiveness of stock $i$ to changes in the market mode $I_{0}(t)$, $\alpha_{i0}$ is the excess volatility compared to the market and $c_{i}(t)$ are the residual log volatilities. 

A standard linear regression of $\omega_{i}(t)$ against $I_{0}(t)$ brings to the surface the residual volatilities $c_{i}(t)$ that the market as a whole cannot explain. The matrix of standardised $c_{i}(t)$ for all stocks is labeled $\bm{X}^{(\text{market})}$. We call $\bm{G}$ the correlation matrix of $\bm{X}^{(\text{market})}$, with entries
\begin{equation}
G_{ij}=\frac{1}{T}\sum_{t=1}^{T}c_{i}(t)c_{j}(t) \ . \label{GEqn}
\end{equation}
By definition, the matrix $\bm{G}$ will have the influence of the market mode removed through Eq. \eqref{MarketEqn}. This cleaning procedure also makes the correlation structure more stable \cite{Borghesi2007}, and therefore we will be working with the matrix $\bm G$ from now on. 

As we did with the matrix $\bm{E}$, we again fit the Mar\v{c}enko-Pastur (MP) distribution -- this time to the empirical eigenvalue distribution of $\bm{G}$. This is justified even in the presence of autocorrelations since in the bulk the amount of memory is quite low. We can see this empirically by computing the median $L_{cut}$ over principal components that have eigenvalues below $\lambda_{+}$ for the fitted MP distribution, which is $2$. The values are quite close to $1$, which is the value we would find for white noise. In the presence of weak autocorrelations, \cite{Burda2005} showed that the distribution of eigenvalues in the bulk differs slightly to the MP distribution. We clearly see this distortion in our Fig. \ref{fig:EigenvalueSpec_detrend}, which bears some similarity in shape with the pdf calculated and plotted in \cite{Burda2005} (see Fig. 1 there). However, the MP distribution is a simpler and very good approximation, especially for the edge points in Fig. \ref{fig:EigenvalueSpec_detrend}. We expect that the number of eigenvalues beyond the bulk should increase, since the removal of the market mode makes the true correlation structure more evident and lowly intra-correlated clusters more visible \cite{Borghesi2007}. This is confirmed by the results that are detailed in Fig. \ref{fig:EigenvalueSpec_detrend}, where we see that the number of eigenvalues beyond the bulk (shown in the inset plot in Fig.\ref{fig:EigenvalueSpec_detrend}) has indeed increased from $22$ to $35$. Note that we also see from Fig. \ref{fig:MarchenkoPastur} that the best fit $q$ and $\sigma$ for $\bm{E}$ and $\bm{G}$ are quite similar, which matches the theoretical result of \cite{Bloemendal2016}.

With this finding in hand, we can safely disregard all principal components corresponding to eigenvalues within the MP sea. This observation already drastically reduces the maximum value of eligible components -- which we call $m_{\mathrm{max}}$ -- from $1202$ to $35$ for the empirical data in \ref{DataCleaning}. We also recall that the eigenvectors of $\bm{G}$ have an economic interpretation according to the Industrial Classification Benchmark (ICB) supersectors -- for more details, see \ref{PortfolioOpt}.  
 
\subsection{Regression of principal components} \label{Regression}
Considering the matrix $\bm{G}$ in Eq. \eqref{GEqn}, where the influence of the market mode has been removed, we must now assess how each stock $i$'s log-volatility is related to the log-volatility of each principal component. We achieve this result by regressing $c_{i}(t)$ in Eq. \eqref{MarketEqn} against the average behaviour of the log-volatility for the principal components. The average behaviour $I_{p}(t)$ is defined as 
\begin{equation}
I_{p}(t)=\sum_{i=1}^{N}w_{ip}c_{i}(t) \ ,\qquad p=1,...,m_{\mathrm{max}}\ ,\label{PortfolioMode}
\end{equation}
i.e. it is the weighted average log-volatility of the $p$-th principal component, where $w_{ip}$ is the $i$-th entry of the $p$-th eigenvector of $\bm G$. Eq. \eqref{PortfolioMode} is therefore the projection of the residue $c_{i}(t)$ onto the $m_{\mathrm{max}}$ principal components.  

The principal components are an orthogonal basis for the correlation matrix $\bm{G}$, and represent important features that determine fluctuations in the $c_{i}$'s. Therefore, it makes sense to define a factor model -- which we call the ``$m_{\mathrm{max}}$-based PCA factor model" -- where the explanatory variables are the $m_{\mathrm{max}}$ principal components \cite{Jolliffe2002,Diana2002}
\begin{equation}
c_{i}(t)=\sum_{p=1}^{m_{\mathrm{max}}}\beta_{ip}I_{p}(t)+\epsilon_{i}(t) \ , \label{FactorModel}
\end{equation}      
where $\epsilon_{i}(t)$ is a white noise term with zero mean and finite variance, and $c_{i}(t)$ are the residual volatilities defined in Eq. \eqref{MarketEqn}. Here, $\beta_{ip}$ is the responsiveness of $c_{i}(t)$ to changes in $I_p(t)$, indicating whether the log-volatility of stock $i$ is higher ($\beta_{ip}>1$) or lower than $I_{p}(t)$ ($\beta_{ip}<1$). 

We can now find $\beta_{ip}$ by regressing the previously obtained input $c_{i}(t)$ against all the $I_{p}$'s. This will separate the signal explained by the principal components from the residual noise present in the system. The regression will be performed using a lasso method (see \ref{Lasso} for details).

\subsection{Assessing memory contribution} \label{MemoryContribution}
The next step of our methodology consists in estimating the memory contribution of the $m=1,2,...,m_{\mathrm{max}}$ components. 

Fixing $m$, we compute for each stock the quantity
\begin{equation}
d_{i}^{(m)}(t)=c_{i}(t)-\sum_{p=1}^{m}\beta_{ip}I_{p}(t) \ ,  \ \ i=1,...,N\ . \label{ResidueDfn}
\end{equation}
Here, the $\beta_{ip}$ are the coefficients obtained with the regression in Eq. \eqref{FactorModel}. The $d_i^{(m)}(t)$ are the residues after the removal of the first $m$ components.  

\begin{figure}
\begin{subfigure}{0.475\textwidth}
\centering
\includegraphics[width=\textwidth]{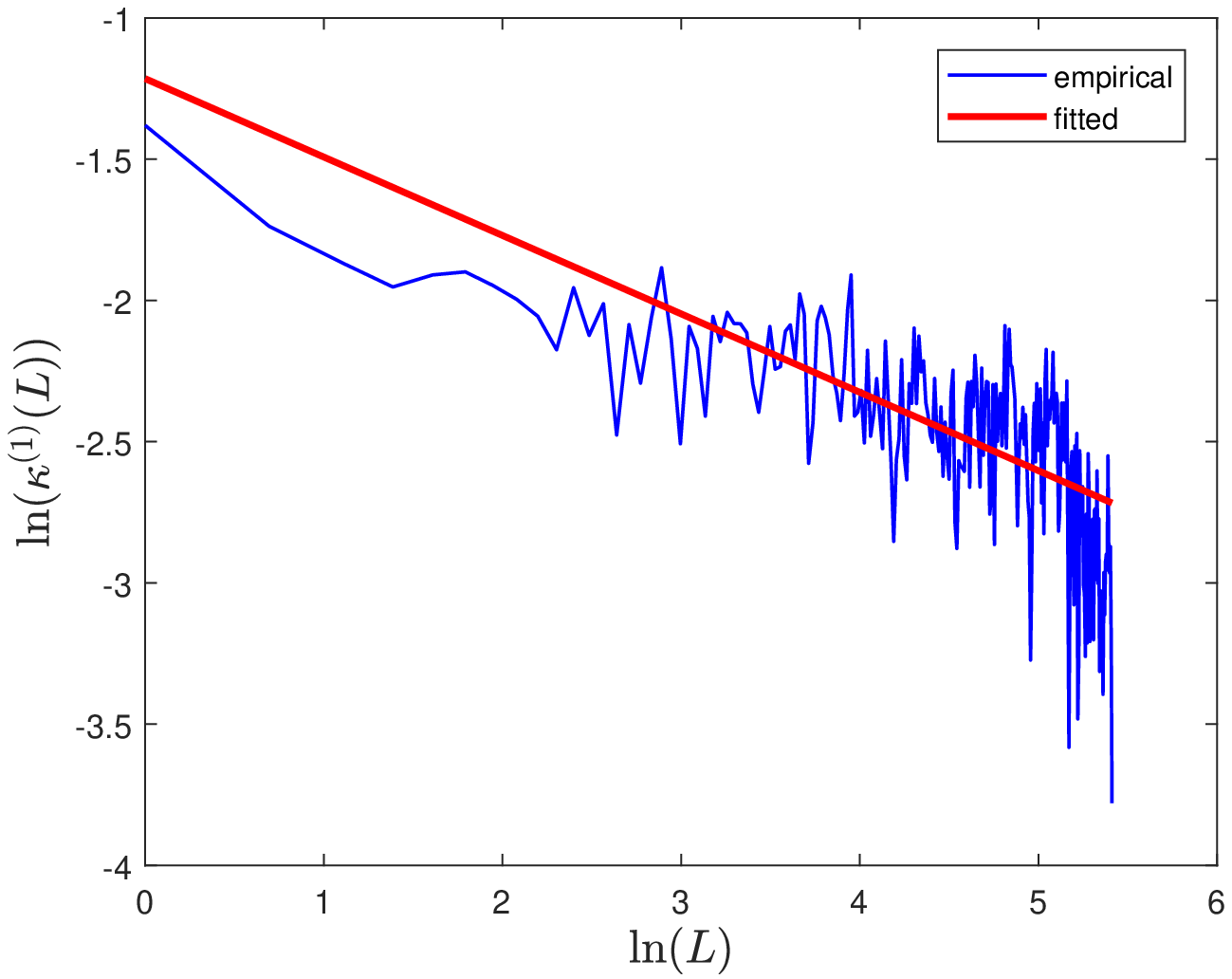}
\caption{\label{fig:AutoCorr_st49_1comp}$\kappa^{(1)}(L)$}
\end{subfigure}
\begin{subfigure}{0.475\textwidth}
\centering
\includegraphics[width=\textwidth]{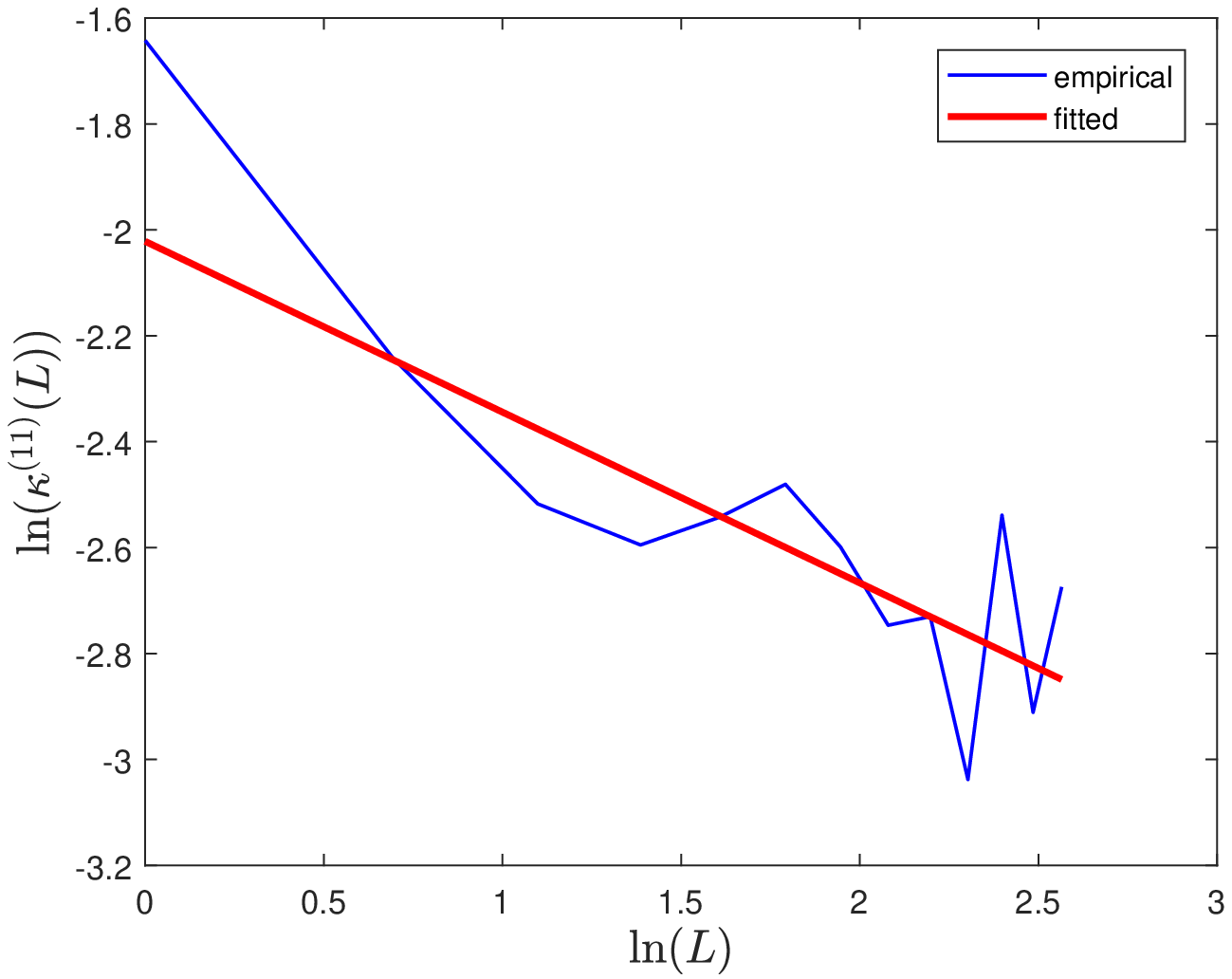}
\caption{\label{fig:AutoCorr_st49_11comp}$\kappa^{(11)}(L)$}
\end{subfigure}
\caption{Plots in log-log scale of $\kappa^{(m)}(L)$ in blue, which is given in Eq. \eqref{ACF} with $x(t)=d_{i}^{(m)}(t)$, for the stock ALJ Regional Holdings (ALJJ). Here, $m=1$ on the left and $m=11$ on the right. In red the lines of best fit (using the Theil Sen estimator \cite{Theil1958}), which gives the power law decay exponent $\beta^{\text{vol}}$ as $0.277$ and $0.322$ for the left and right plot respectively.}
\label{fig:AutoCorr_st49_comp}
\end{figure}

Using the $d_{i}^{(m)}(t)$, we can first compute their temporal autocorrelation $\kappa_i^{(m)}(L)$ in Eq. \eqref{ACF} for different values of the lag $L$ between $L=1$ and $L=T-1$. We generically find that the $\kappa_{i}^{(m)}(L)$ follow a power law decay as a function of $L$ -- see examples in Fig. \ref{fig:AutoCorr_st49_comp} depicting the $\kappa_{i}^{(m)}(L)$ for $m=1,11$ for ALJ Regional Holdings (ALJJ), a stock included in our empirical dataset in \ref{DataCleaning}. As more components are removed (i.e. as $m$ is increased), the exponent $\beta^{\text{vol}}$ defined in Eq. \eqref{PowerLawACF} for ALJJ and plotted in Fig. \ref{fig:AutoCorr_st49_comp} increases from $0.277$ to $0.322$. This result is what one would expect since the amount of memory accounted for will decrease as more components are removed.

Numerically integrating the $\kappa_{i}^{(m)}(L)$, we obtain a set of integrated memory proxies $\eta_{i}^{(m)}$ (see Eq. \eqref{IntProxy}), one for each asset $i$ and for each number $m$ of removed components.  In general, the $\eta_{i}^{(m)}$ are non-increasing functions of $m$, since the further removal of subsequent components is bound to decrease the residual memory level 
present in the system.

We now define
\begin{equation}
\zeta(m)=\mathrm{median}\left(\frac{\eta_{i}^{(m)}}{\eta_{i}^{(0)}}\right) \ , \label{MemoryReductionEqn}
\end{equation}
where $\eta_{i}^{(m)}$ are the integrated proxies, and $\eta_{i}^{(0)}$ are just the integrated proxies of the residual volatilities $c_{i}(t)$ defined in Eq. \eqref{MarketEqn}.  $\zeta(m)$ thus represents the ``average" behaviour over all stocks of how much each of the principal components contributes to the memory. It is again a non-increasing function of $m$, and by definition $\zeta(m)<1$ for all $m$.

\begin{figure}
\begin{subfigure}{0.475\textwidth}
\centering
\includegraphics[width=\textwidth]{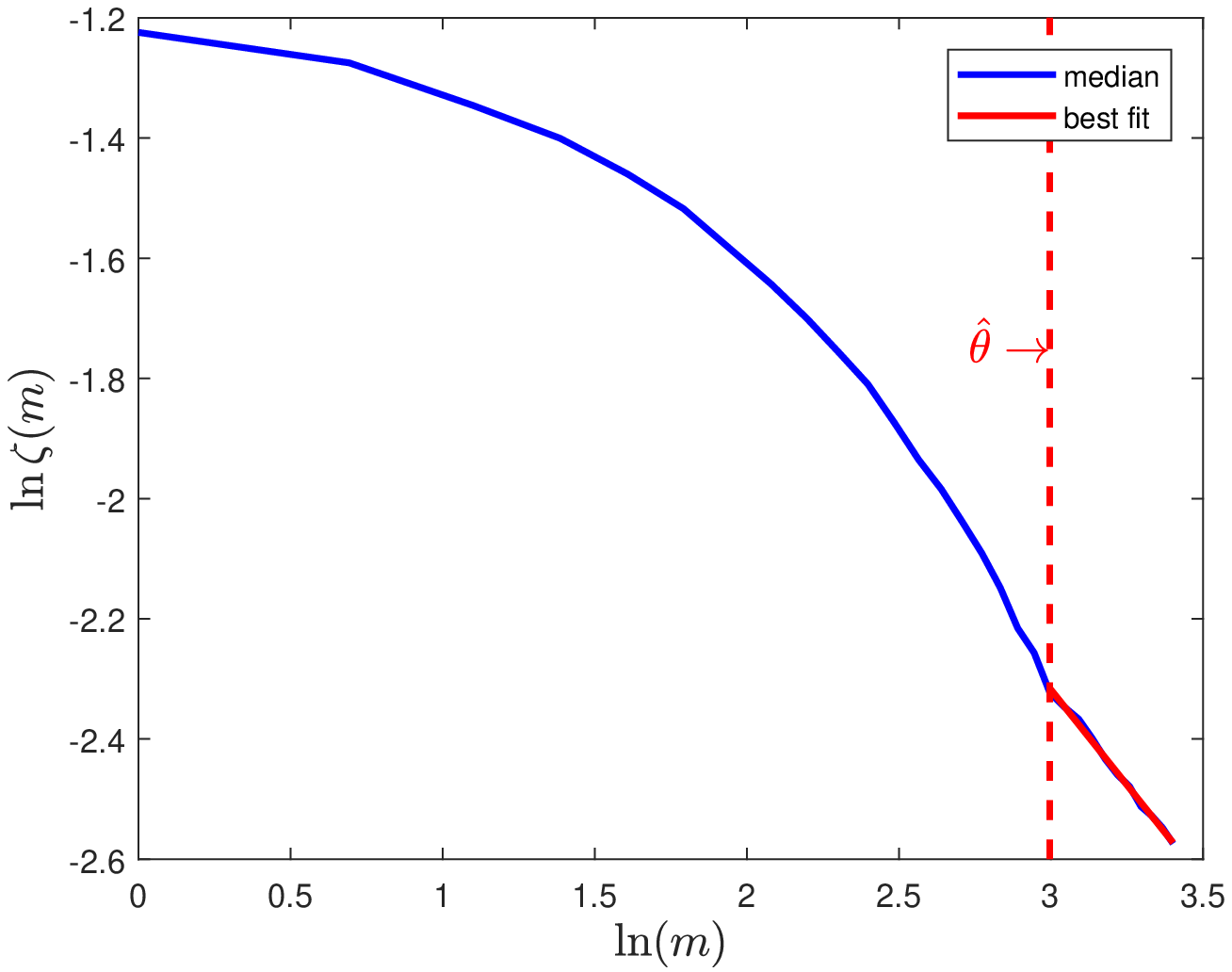}
\caption{
\label{fig:loglog_samples_medianratio}homogeneous}
\end{subfigure}
\begin{subfigure}{0.475\textwidth}
\centering
\includegraphics[width=\textwidth]{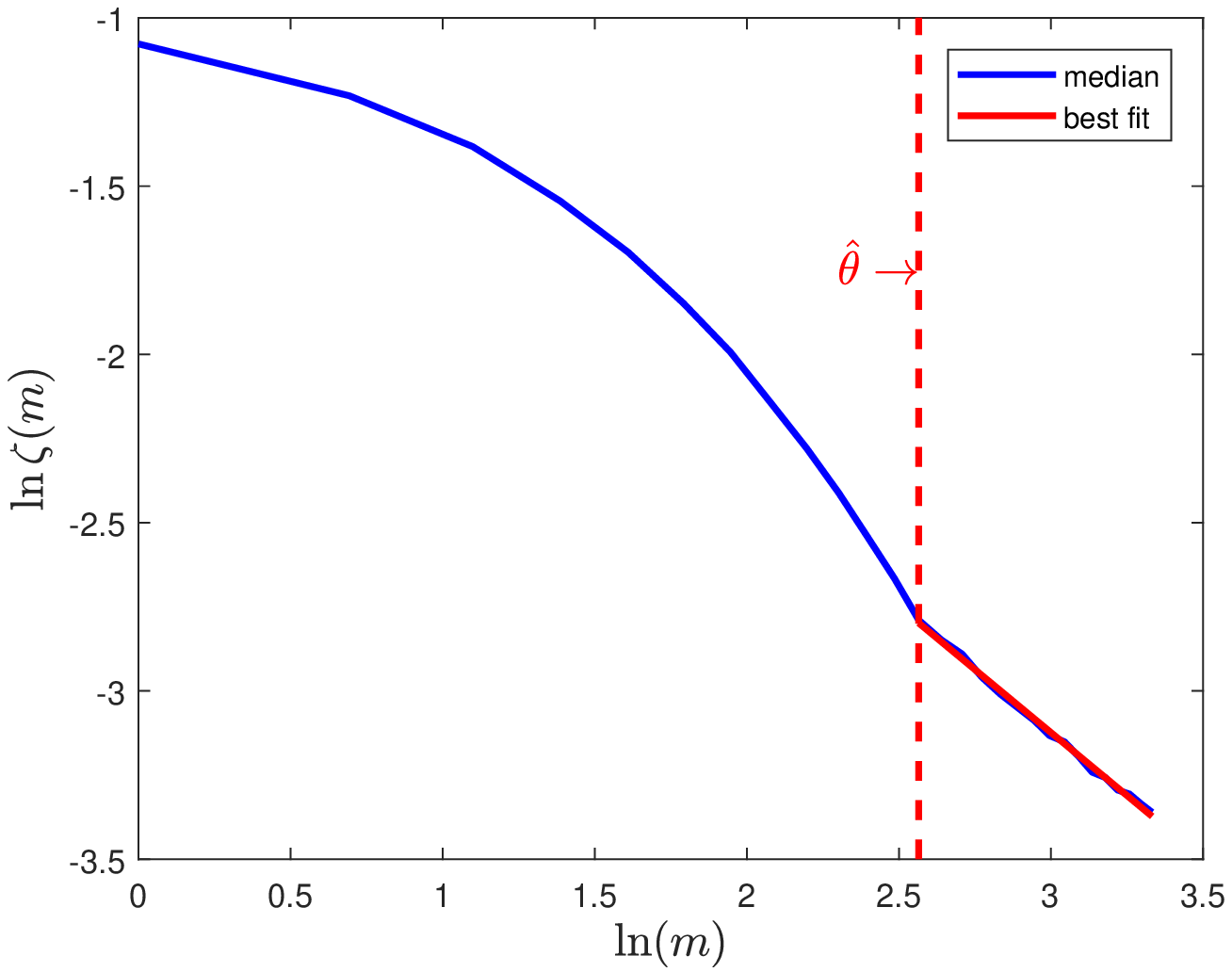}
\caption{\label{fig:loglog_samples_medianratio_het}heterogeneous}
\end{subfigure}
\\
\begin{subfigure}{0.475\textwidth}
\centering
\includegraphics[width=\textwidth]{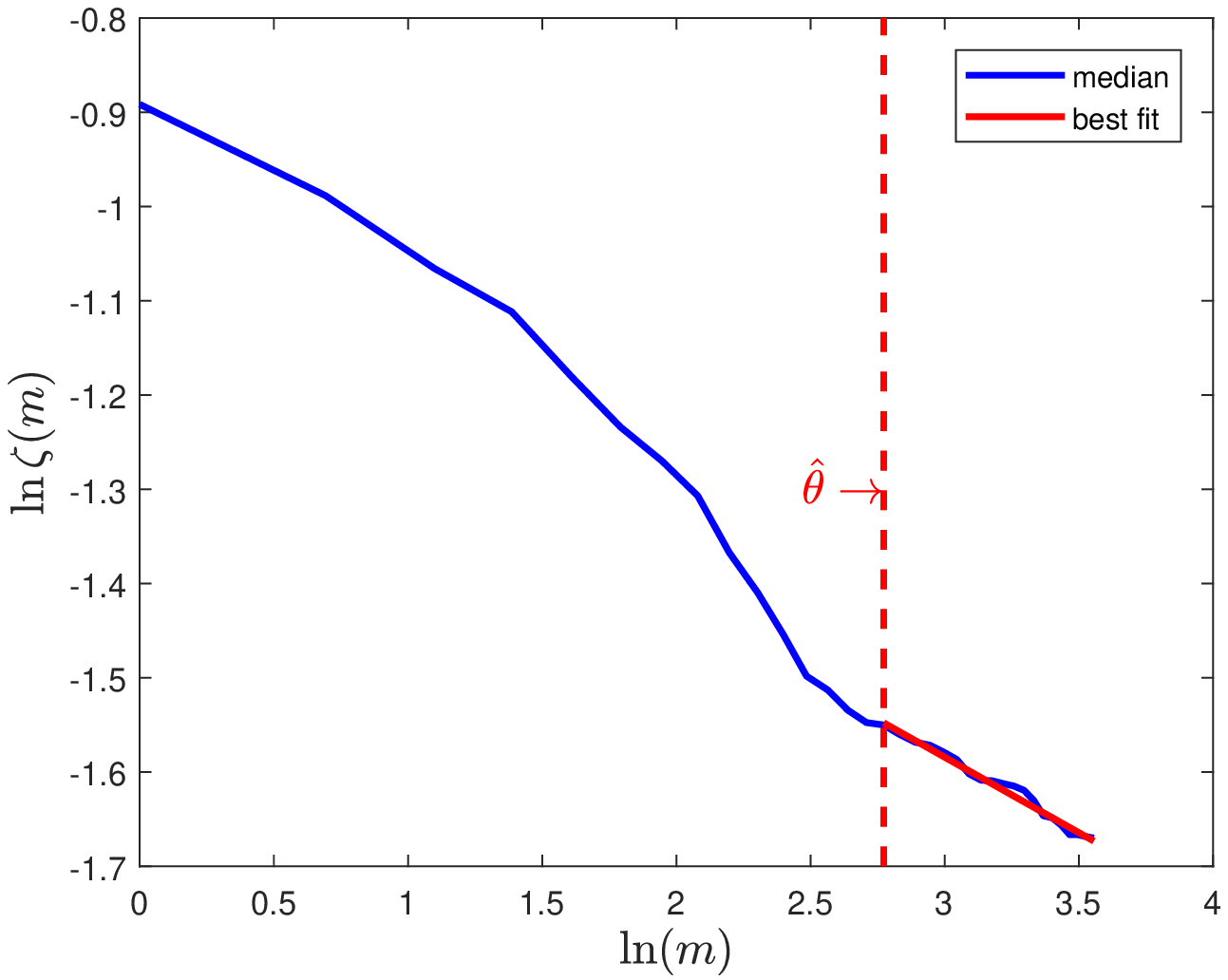}
\caption{\label{fig:loglog_hattheta}empirical}
\end{subfigure}
\caption{(Top left) Plot of $\ln(\zeta(m))$ vs $\ln(m)$ for the homogeneous synthetic system defined in Section \ref{Synthetic}. The blue line is the value of $\zeta(m)$ across all assets, with the dashed red line indicating $\hat{\theta}=20$, the point at which the concavity changes. (Top right) Same plot but for the heterogeneous simulated system described in Section \ref{Synthetic}, where $\hat{\theta}=13$. (Bottom) Same plot but for the empirical dataset described in \ref{DataCleaning}, yielding $\hat{\theta}=16$. These values of $\hat{\theta}$ imply that the number $m^\star$ of principal components to retain should be $m^\star=19,12,15$ respectively.}
\label{fig:mstar_comparison}
\end{figure}

In Fig. \ref{fig:loglog_hattheta}, we plot $\zeta(m)$ in log-log scale for both the empirical and synthetic datasets in \ref{DataCleaning} and Section \ref{Synthetic} respectively. We observe a striking change in concavity at some value $\theta$, which we interpret as follows: to the right of $\theta$, the amount of memory left unexplained in the system changes very slowly when more and more components are progressively included. This clearly signals that we have reached the ``optimal stopping" point $m^\star$ beyond which the inclusion of further components would not add more information. 

Beyond $\theta$, the behaviour of $\zeta(m)$ is power-law 
\begin{equation}
\zeta(m) \sim m^{-\gamma} \qquad m \geq \theta \ , \label{ThetaPowerLaw}
\end{equation}
where $\gamma$ is the exponent. Using the fitting procedure for $\theta$ described in \ref{ThetaHatFitting} produces the optimal integer estimator $\hat{\theta}$. Since the value of $\hat{\theta}$ indicates that for $m<\hat{\theta}-1$, $\zeta(m)$ decreases more rapidly than a power law, we can safely set
\begin{equation}
m^{\star}=\hat{\theta}-1 \ .
\end{equation}

\subsection{Summary of the procedure} \label{ProcedureSummary}
The procedure to select the optimal number $m^{\star}$ of principal components to retain is summarised here for a general, standardised data-matrix $\bm{X}$ containing long memory effects (justifications for the steps can be found in the Sections labelled in brackets after each step):
\begin{enumerate}
\item Remove any global effect from $\bm{X}$ to form $\bm{X}^{(\text{market})}$, whose entries are the residues $c_{i}(t)$ defined in Eq. \eqref{MarketEqn} [Section \ref{DetrendMarketMode}]. 
\item Compute the correlation matrix $\bm{G}$ of $\bm{X}^{(\text{market})}$ and find the empirical probability density of its eigenvalues. Find the number of eigenvalues $m_{\mathrm{max}}$ exceeding $\lambda_{+}$, the upper edge of the Mar\v{c}enko-Pastur distribution in Eq. \eqref{MPDist} [Section \ref{MarchenkoPastur}].
\item Forming the $m_{\mathrm{max}}$-based PCA factor model from Eq. \eqref{FactorModel}, use lasso regression (see \ref{Lasso}) to find the set of parameters $\beta_{ip}$. This is achieved by regressing the residues $c_{i}(t)$ against the average behaviour of principal components $I_{p}(t)$ $p=1,...,m_{\mathrm{max}}$ [Section \ref{Regression}].
\item Using these $\beta_{ip}$'s, determine for each $m=1,...,m_{\mathrm{max}}$ and stock $i$ the residue $d_{i}^{(m)}(t)$ given in Eq. \eqref{ResidueDfn} [Section \ref{MemoryContribution}].
\item From the $d_{i}^{(m)}(t)$, compute the temporal autocorrelations $\kappa_i^{(m)}(L)$ for different values of $L$, and by integration determine the proxies $\eta_i^{(m)}$. Construct $\zeta(m)$ from Eq. \eqref{MemoryReductionEqn} [Section \ref{MemoryContribution}]. 
\item Use the fitting procedure in \ref{ThetaHatFitting} to find $\hat{\theta}$, the best estimator of $\theta$ -- the point at which the concavity of $\zeta(m)$ changes -- defined in Eq. \eqref{ThetaPowerLaw}. Finally, the optimal number of principal components to retain is $m^{\star}=\hat{\theta}-1$ [Section \ref{MemoryContribution}].
\end{enumerate}    

\section{Applying our method to synthetic and empirical data} \label{ApplyingMethod}

In this Section, we test our method on synthetically generated data and on an empirical data set defined in \ref{DataCleaning}. 

\subsection{Synthetic System Setup} \label{Synthetic}

A paradigmatic example of stochastic process with long memory is the Fractional Gaussian Noise (FGN). The FGN with Hurst exponent $H$ is the process $Y(t)$ with an autocorrelation function \cite{Beran2017} given by
\begin{equation}
\kappa_{FGN}(L)=\frac{1}{2}\left(|L-1|^{2H}-2|L|^{2H}+|L+1|^{2H}\right) \sim H(2H-1)L^{2H-2} \ . \label{fGn_ACF}
\end{equation}
Eq. \eqref{fGn_ACF} indeed shows that the FGN has long memory since its autocorrelation function follows a power law decay as described in Section \ref{VolatilityClustering}. In particular, for $1/2<H<1$ ($H=1/2$ corresponds to the standard white noise) we have a process with positive autocorrelation, a feature that is shared by financial data \cite{Cont2001}. Increasing $H$ will enhance the strength of the memory present in the FGN since $\kappa_{FGN}(L)$ will decay more slowly in this case. We shall use the method detailed in \cite{Dietrich1997} to generate realisations of FGN.  

For our synthetic setting, we consider a fictitious market made of $N$ stocks, and simulate the log-volatility $\omega_i(t)$ of each stock over a time-window $T$. To this end, we make use of the widely recognised fact that empirical data from finance are often organised into clusters \cite{Mantegna1999,Musmeci2015,Musmeci2015b,Musmeci2016}. We therefore impose that the stocks are organised into $K$ disjoint clusters, each containing $N_{k}$ stocks. 

Next, we generate a fictitious market mode $I_0(t)$ that affects all stocks \cite{Laloux1999,Plerou2002,Singh2016}. This is simply a FGN process with Hurst exponent $H_0$, which we will set to $0.9$ in our simulations. We also fix the variance of $I_{0}(t)$ to be $1$.

Our simulated log-volatility processes will thus read
\begin{equation}
\omega_i(t)=\beta_0 I_0(t) +\beta_{k(i)}I_{k(i)}(t)+\epsilon_i(t)\ ,\label{volproc}
\end{equation}
where $k(i)$ denotes the index of the cluster the asset $i$ belongs to, and the $I_k(t)$'s are FGN processes with Hurst exponents $H_k$ and fixed variance of $1$. The $\epsilon_{i}$'s are  white noise terms with zero mean and variance $\phi$. 

Typical values we use in our simulations are $N=1200$, $T=4000$ and $K=30$. We simulate two markets with different internal arrangements of the clusters: the first one is homogeneous, where the size of each cluster is exactly $40$, and the second is heterogeneous, meaning that each cluster has a different number of stocks. The latter case is particularly significant since the cluster sizes present in financial data as well as in many other systems are known to be heterogeneous \cite{Mantegna1999,Musmeci2015}. To generate the heterogeneous system, we use the procedure described in \cite{Lancichinetti2008}, which yields power-law distributed cluster sizes, a key property of real world data \cite{Palla2005}. The particular realisation of this method that we use to generate cluster sizes for $N=1200$ has a mean number of stocks in each cluster of $40$ and a standard deviation of $26.2$. We also set $\beta_{0}=1.3$, while $\beta_{k}$ are values between $0.14$ and $1$, and $H_{k}$ is an equally spaced sequence between $0.7$ and $0.9$. This choice ensures that clusters with a higher $\beta_{k}$ will also have a higher $H_{k}$, to make contact with the empirical result of \cite{Micciche2013} that stocks with higher volatility cross-correlation have a longer memory. Finally, $\phi$ is fixed to be $1$, the same as the variance of the time series $I_{0}(t)$ and $I_{k}(t)$. Note that we also simulate the same system using instead an Autoregressive process of $1$ lag (AR(1)) \cite{Box2015} in \ref{Autoregressive}, where we show that our method can be applied in this case too, but is less accurate. This supports our reasoning that that slow decrease to the right of $\hat{\theta}$ in Fig. \ref{fig:loglog_hattheta} is more applicable to long memory processes versus short ones.

Arranging the log-volatilities $\omega_i(t)$ in a rectangular data-matrix $\bm X$, we can then feed $\bm X$ into our algorithmic procedure and check how many significant components $m^\star$ it retrieves. A desirable feature of our synthetic model is that it is rather easy to estimate \emph{a priori} how many eigenvalues of $\bm E=(1/T)\bm X^\dagger \bm X$ (or rather of its de-trended counterpart $\bm G$) contain information that can be separated from pure noise. This occurs because each cluster corresponds to a principal component and hence the number of eigenvalues beyond the bulk is just $K$. This makes the \emph{a posteriori} comparison all the more interesting.

\subsection{Results for synthetic and empirical data} \label{Synthetic_results}
\begin{figure}
\begin{subfigure}{0.475\textwidth}
\centering
\includegraphics[width=\textwidth]{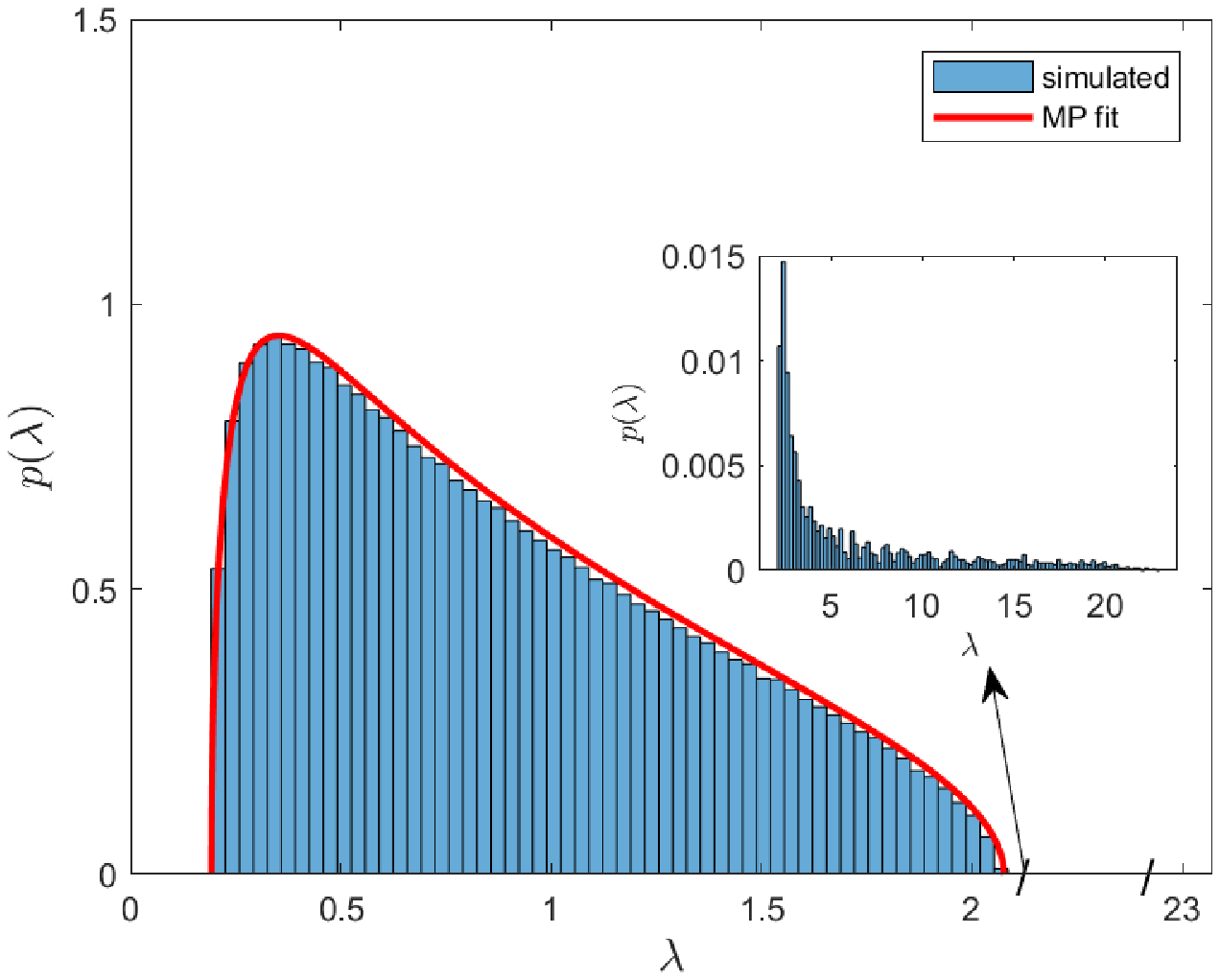}
\caption{homogeneous}
\label{fig:EigenvalueSpec_simulated} 
\end{subfigure}
\begin{subfigure}{0.475\textwidth}
\centering
\includegraphics[width=\textwidth]{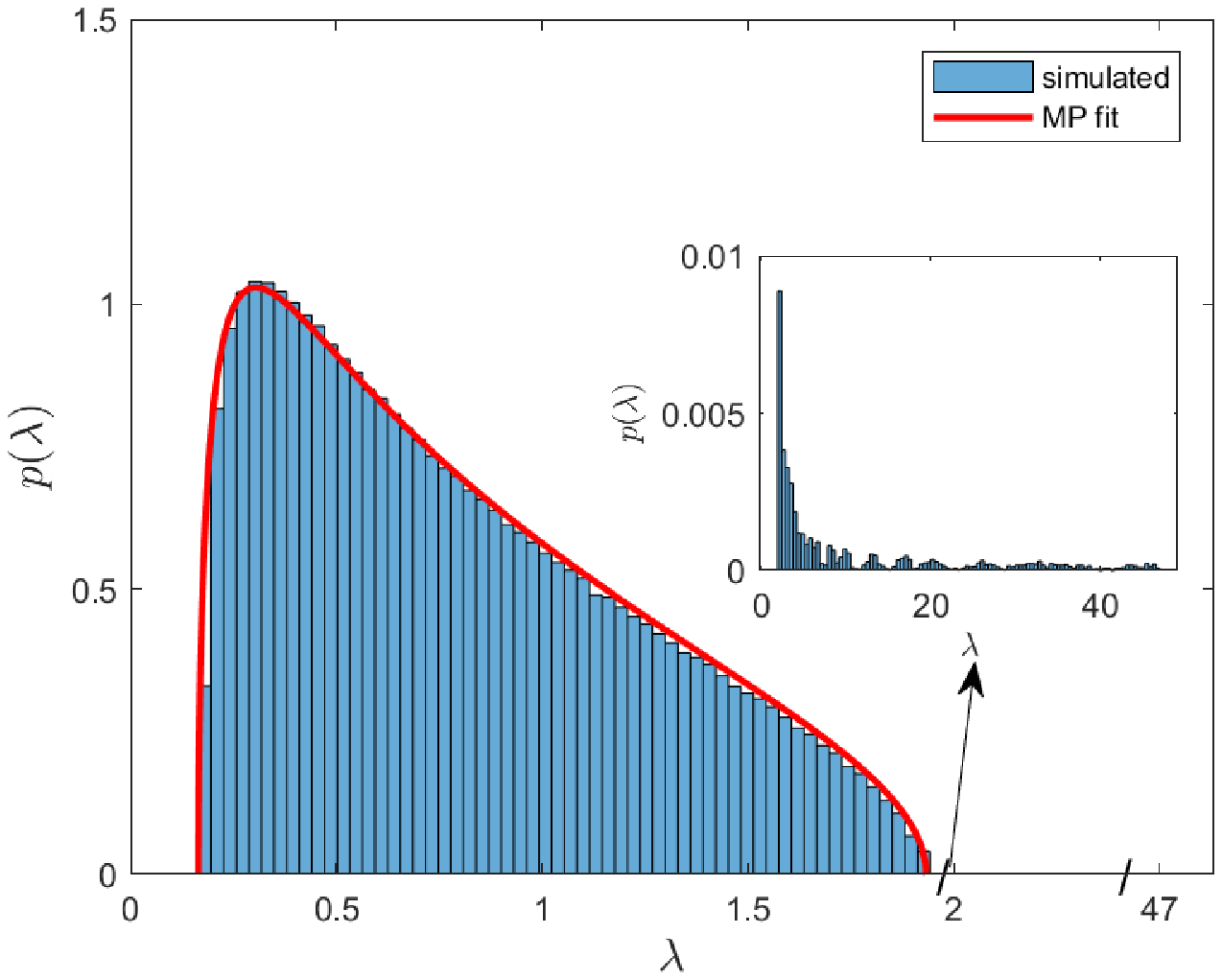}
\caption{heterogeneous}
\label{fig:EigenvalueSpec_simulated_het} 
\end{subfigure}
\caption{Histograms of eigenvalues of the matrix $\bm G$ for $100$ samples of the synthetic market with $N=1200$, $T=4000$, and $K=30$ clusters.
The values of the $\beta$ coefficients and of the Hurst exponents are as in the main text. 
(Left) Homogeneous system with $40$ stocks in each cluster. In red the best fit Mar\v{c}enko-Pastur distribution of Eq. \eqref{MPDist} with parameters $q=0.284\pm 0.002$ and $\sigma=0.939\pm 0.001$ with upper edge $\lambda_{+}=2.0756$. The inset includes the $m_{\mathrm{max}}=30$ eigenvalues beyond $\lambda_{+}$.  (Right) Same plot but for a heterogeneous system with the same parameters, but a different cluster structure defined in Section \ref{Synthetic}. Here $\lambda_{+}=1.9322$, $q=0.299\pm 0.004$, and $\sigma=0.898\pm 0.002$. Finally, in this case there are $m_{\mathrm{max}}=28$ eigenvalues beyond $\lambda_{+}$.}
\label{fig:synthetic_system}
\end{figure}
We simulate $100$ independent samples of our synthetic market, after checking that the statistics was sufficient to be confident on the stability of our results, and we follow the procedure set out at the end of Section \ref{MemoryContribution} to select $m^{\star}$. First, we checked the eigenvalue distributions of the correlation matrix obtained from the simulated $\bm X$. We see from Fig. \ref{fig:synthetic_system}, which are histograms of the bulk eigenvalues of $\bm G$ for all samples for the homogeneous (left) and heterogeneous (right) systems respectively, that the bulk of the eigenvalues is well fitted by the Mar\v{c}enko-Pastur distribution in red. There are $m_{\mathrm{max}}=30$ (homogeneous) and $m_{\mathrm{max}}=28$ (heterogeneous) eigenvalues beyond the bulk (depicted in the insets) that carry genuine information. This again shows that in the synthetic case the autocorrelations are also weak. We see this again by calculating the median $L_{cut}$ of Eq. \eqref{IntProxy} for the synthetic systems to be $2$ -- again close to $1$, which is what we would expect for white noise, hence we can still use the MP distribution as an approximation. We also remark that the MP fits in Figs. \ref{fig:EigenvalueSpec_simulated} and \ref{fig:EigenvalueSpec_simulated_het} are better than that of Figs. \ref{fig:EigenvalueSpec} and \ref{fig:EigenvalueSpec_detrend} because we can tune the white noise in our synthetic data so that the bulk in this region behaves more similar to white noise. This is achieved by changing the value of $\phi$.

For each sample we find the median $\zeta(m)$, plotting it in log-log scale in Fig. \ref{fig:loglog_samples_medianratio} for the homogenous system and in Fig. \ref{fig:loglog_samples_medianratio_het} for the heterogeneous one. 

As already described, the optimal value $m^{\star}$ turns out to be $19$ and $12$  for the homogenous and heterogeneous systems respectively. The fact that $m^{\star}$ is lower for the heterogeneous system makes sense since its broad, power law distributed values of $N_{k}$ mean that more of the memory of the system is contained in earlier principal components, whose $N_{k}$ are larger. Since more of the memory is concentrated in fewer principal components, it is natural that the corresponding values of $m^{\star}$ will be lower for the heterogenous system. On the other hand for the homogenous system, we have that the $N_{k}$ are equal for all $k$, so we can expect that the memory is more evenly distributed across the principal components i.e. that $m^{\star}$ will be larger. We also apply the method in Section \ref{Method} to the data matrix $\bm{X}$ corresponding to the empirical dataset described in \ref{DataCleaning}, for which $m_{\mathrm{max}}=35$ (see caption of Fig. \ref{fig:loglog_hattheta} for details).

\section{Comparison with other heuristic methods to select $m^{\star}$} \label{Comparison}
In this Section, we shall compare our new method with available ``stopping rules" in the literature. Many heuristic methods have been proposed in order to determine $m^{\star}$, generally falling into three categories: subjective methods, distribution-based methods and computational procedures \cite{Jackson1993,Jolliffe2002}. We describe here the most common ones in each category.

In the class of subjective methods, we find two similar procedures, the \emph{cumulative percentage of variation} \cite{Sugiyama1976,Huang1992} and \emph{scree plots} \cite{Cattell1966}. The former is based on selecting the minimum value of $m$ such that the cumulative percentage of variation explained by the $m$ principal components exceeds some threshold $\alpha$:
\begin{align}
m^{\star}&=\min_{m}\left\{\Lambda(m) > \alpha\right\} \ , \\
\Lambda(m)&=100\frac{\sum_{p=1}^{m}\lambda_{p}}{N} \ ,
\label{CumVarEqn}
\end{align}
where $\Lambda(m)$ is the $\%$ cutoff, $\alpha$ is the percentage cutoff threshold and $\{\lambda_{p}\}_{p=1}^{m}$ are the first $m$ eigenvalues of $\bm{G}$. Common cutoff ranges lie somewhere between $70\%$ to $90\%$, with a preference towards larger values when it is known or obvious that the first few principal components will explain most of the variability in the data \cite{Jolliffe2002}. An obvious disadvantage of this method is that it relies on the choice of some arbitrary value for the tolerance $\alpha$.

Scree plots involve plotting a `score' representing the amount of variability in the data explained by individual principal components, and then choosing the point at which the plot develops an `elbow', beyond which picking further principal components does not significantly enhance the level of memory already accounted for. This procedure again has the obvious drawback of 
relying on graphical inspection and therefore being even more subjective than the cumulative percentage of variation.

Among the class of distribution-based methods, the most commonly used procedure is the Bartlett Test \cite{Bartlett1950}. This involves testing the null hypothesis \cite{Jolliffe2002}
\begin{equation}
H_{0,m}=\lambda_{m+1}=\lambda_{m+2}=...=\lambda_{N} \ ,
\end{equation}
that is whether the last $N-m$ eigenvalues are identical, against the alternative that at least two of the last $N-m$ eigenvalues are not identical, and repeating this test for various values of $m$. One then selects the maximum value of $m$ for which the outcome of the hypothesis test is significant. Intuitively, this procedure tests whether the last $N-m$ eigenvalues explain roughly the same amount of variability in the data so that they can be regarded as noise, and then takes $m^{\star}$ to be the maximum number of ``significant" eigenvalues. According to this procedure, one first tests $H_{0,N-2}$ i.e. whether $\lambda_{N-1}=\lambda_{N}$. If this hypothesis is not rejected, then one tests $H_{0,N-3}$, and if this is not rejected the exact same test is performed for $H_{0,N-4}$ and so on. The procedure carries on testing each individual $H_{0,m}$ until the first time ($m=m^{\star}-1$) the hypothesis gets rejected at the required confidence level. Since several tests need to be conducted sequentially, the overall significance of the procedure will not be the same as the one imposed for each individual test, with no way of correcting for this bias as the number of tests to be performed is \emph{a priori} unknown. This drawback makes distribution-based methods very impractical with real data \cite{Jolliffe2002}.

The last category (computational procedures) involves the use of cross-validation. Cross-validation requires that some chunks of the original dataset $\bm{X}$ be initially removed. The remaining data matrix entries are used in conjunction with Eq. \eqref{FactorModel} to cast predictions on the removed entries using $m$ principal components. We focus on so called $10$-fold contiguous block cross-validation, which has been argued to be optimal in the sense that it most accurately captures the true structure of the correlation matrix (either $\bm E$ or $\bm G$) \cite{Bun2016}. According to this procedure, we divide the data matrix $\bm{X}$ into $10$ rectangular blocks row-wise, which we call $\bm{X}^{(g)}$ for $g=1,...,10$. For each group $g$, we calculate the correlation matrix $\bm{G}^{(g)}$associated with the matrix $\bm X$ but with the block $\bm{X}^{(g)}$ removed. Next, we take $m$ principal components of $\bm{G}^{(g)}$ and use them in a factor model like in Eq. \eqref{FactorModel} but with $m$ as the upper limit for the sum to predict the values of $\bm{X}^{(g)}$, which we call $\hat{\bm{X}}^{(g,m)}$. We then repeat this procedure for every $m$ and $g$.

After doing so, we can calculate the Prediction Residual Error Square Sum, or $\text{PRESS}(m)$, as a function of $m$. This is the total (un-normalised) squared prediction error for each value and over all blocks
\begin{equation}
\text{PRESS}(m)=\sum_{i=1}^{N}\sum_{g=1}^{10}\sum_{t\in \mathcal{G}_{g}}\left(\hat{\bm{X}}^{(g,m)}_{ti}-\bm{X}^{(g)}_{ti}\right)^{2} \ , \label{PRESS_eqn}
\end{equation}
with $\hat{\bm{X}}^{(g,m)}$ being the matrix of predicted values for block $g$ using $m$ principal components, and $\mathcal{G}_{g}$ indicating the row indices belonging to block $g$. Eq. \eqref{PRESS_eqn} represents the out-of-sample error in predicting the entries of $\bm{X}$, which implies that $\text{PRESS}(m)$ should initially decrease as $m$ increases. However, beyond a certain threshold, $\text{PRESS}(m)$ might start to increase instead, indicating that we are beginning to overfit the data. The optimal $m^\star$ should therefore be chosen to be the value which minimises $\text{PRESS}(m)$, thus striking the optimal balance between increasing the model complexity and overfitting the data. This procedure has an obvious advantage over the previous two categories as it is parameter-free and not subjective. However, one significant drawback for practical purposes is that the procedure becomes computationally very expensive for large datasets due to the typically $\sim \mathcal{O}(N m_{\mathrm{max}})$ regressions that need to be performed from the dataset.

\begin{figure}
\begin{subfigure}{0.475\textwidth}
\centering
\includegraphics[width=\textwidth]{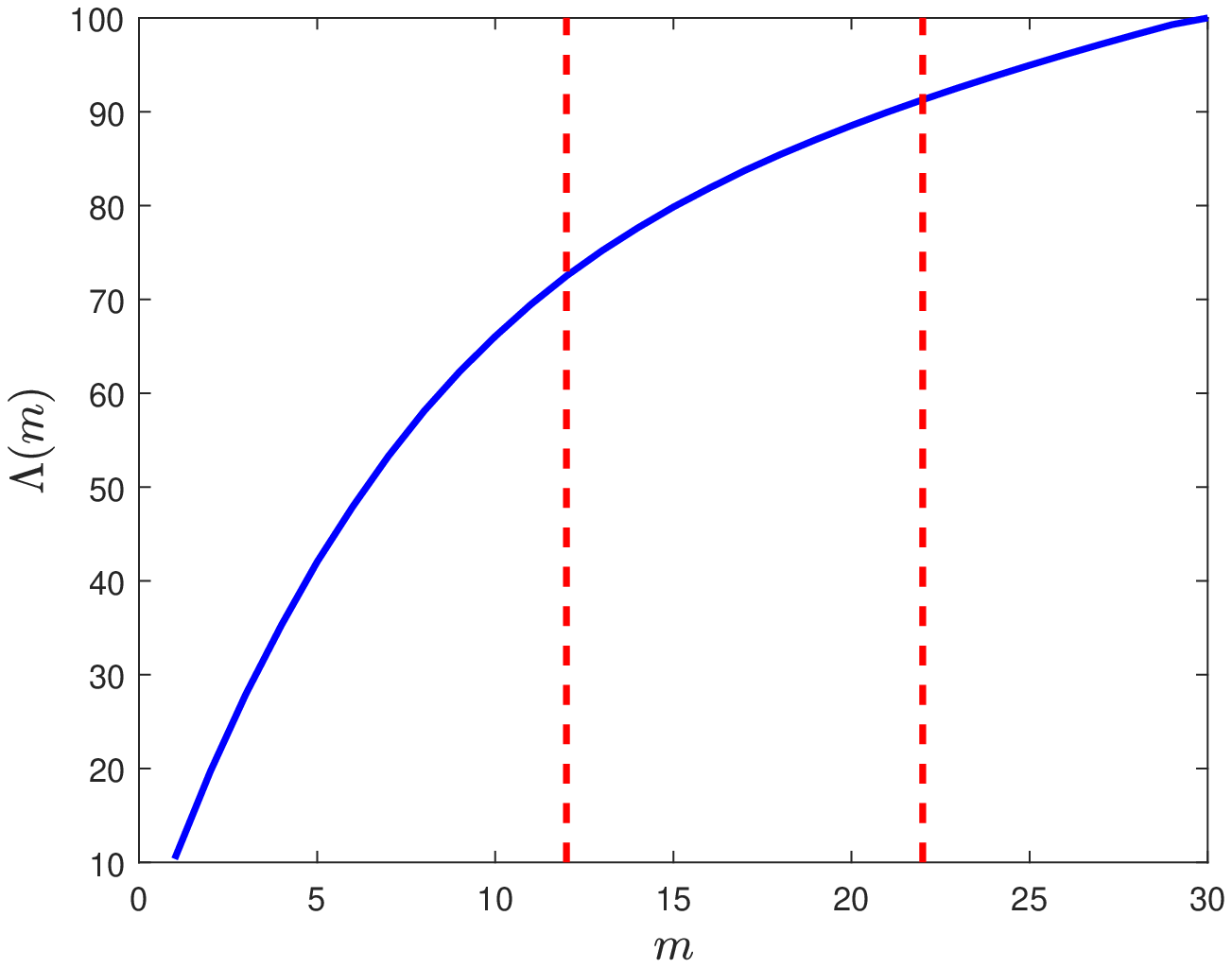}
\caption{\label{fig:cumvar_simulation}homogeneous}
\end{subfigure}
\begin{subfigure}{0.475\textwidth}
\centering
\includegraphics[width=\textwidth]{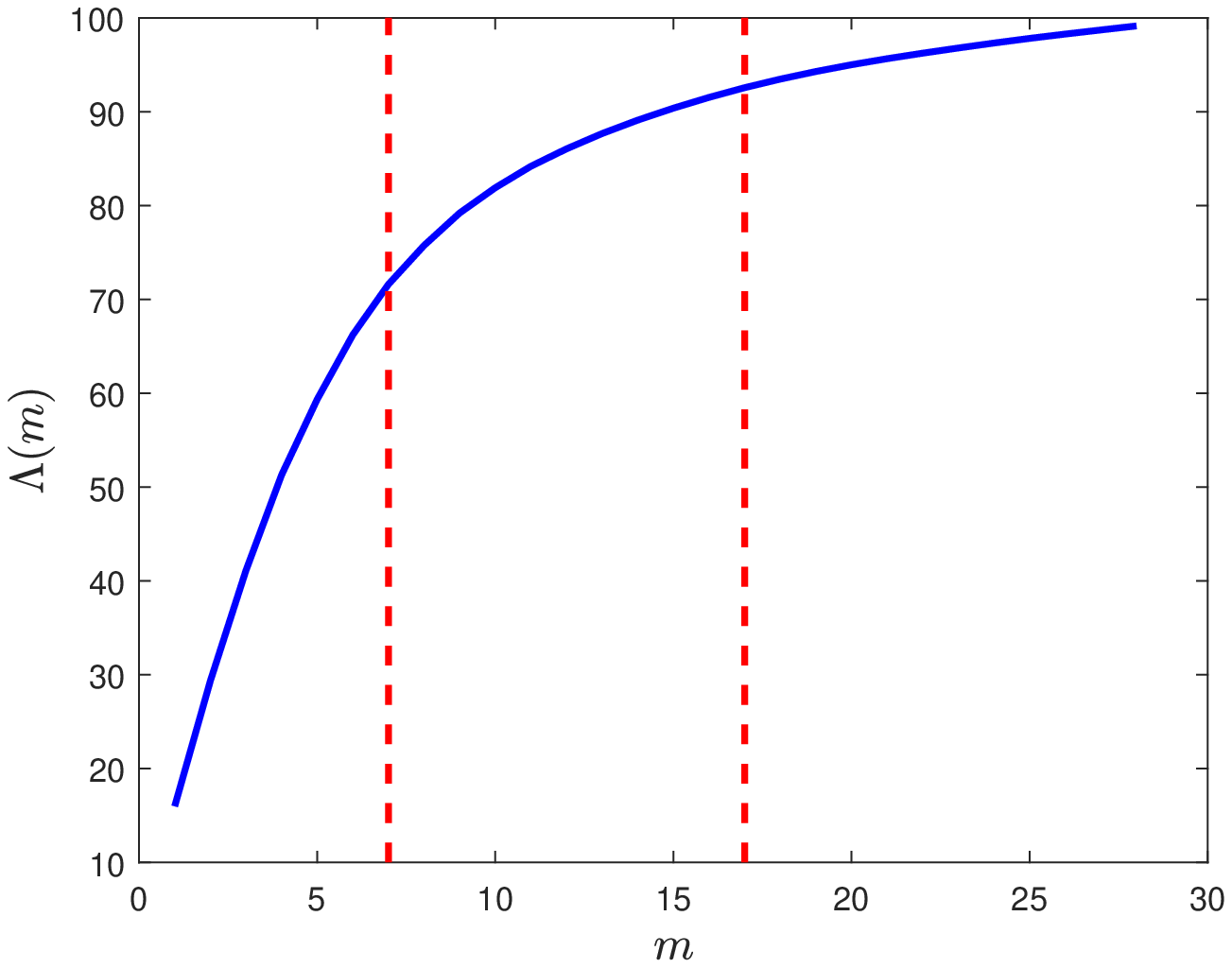}
\caption{\label{fig:cumvar_het}heterogeneous}
\end{subfigure}
\\
\begin{subfigure}{0.475\textwidth}
\centering
\includegraphics[width=\textwidth]{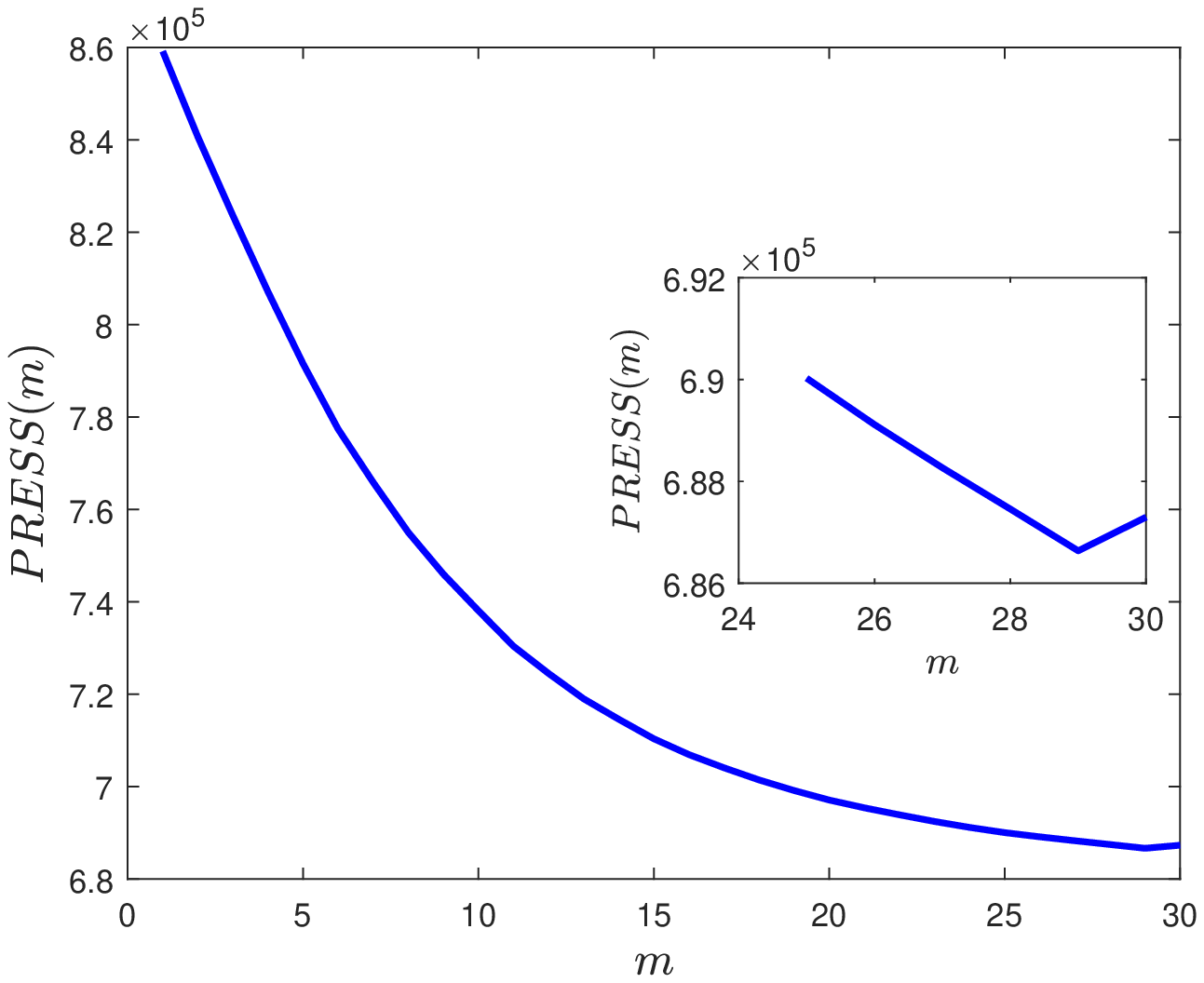}
\caption{\label{fig:crossval_simulation}homogeneous}
\end{subfigure}
\begin{subfigure}{0.475\textwidth}
\centering
\includegraphics[width=\textwidth]{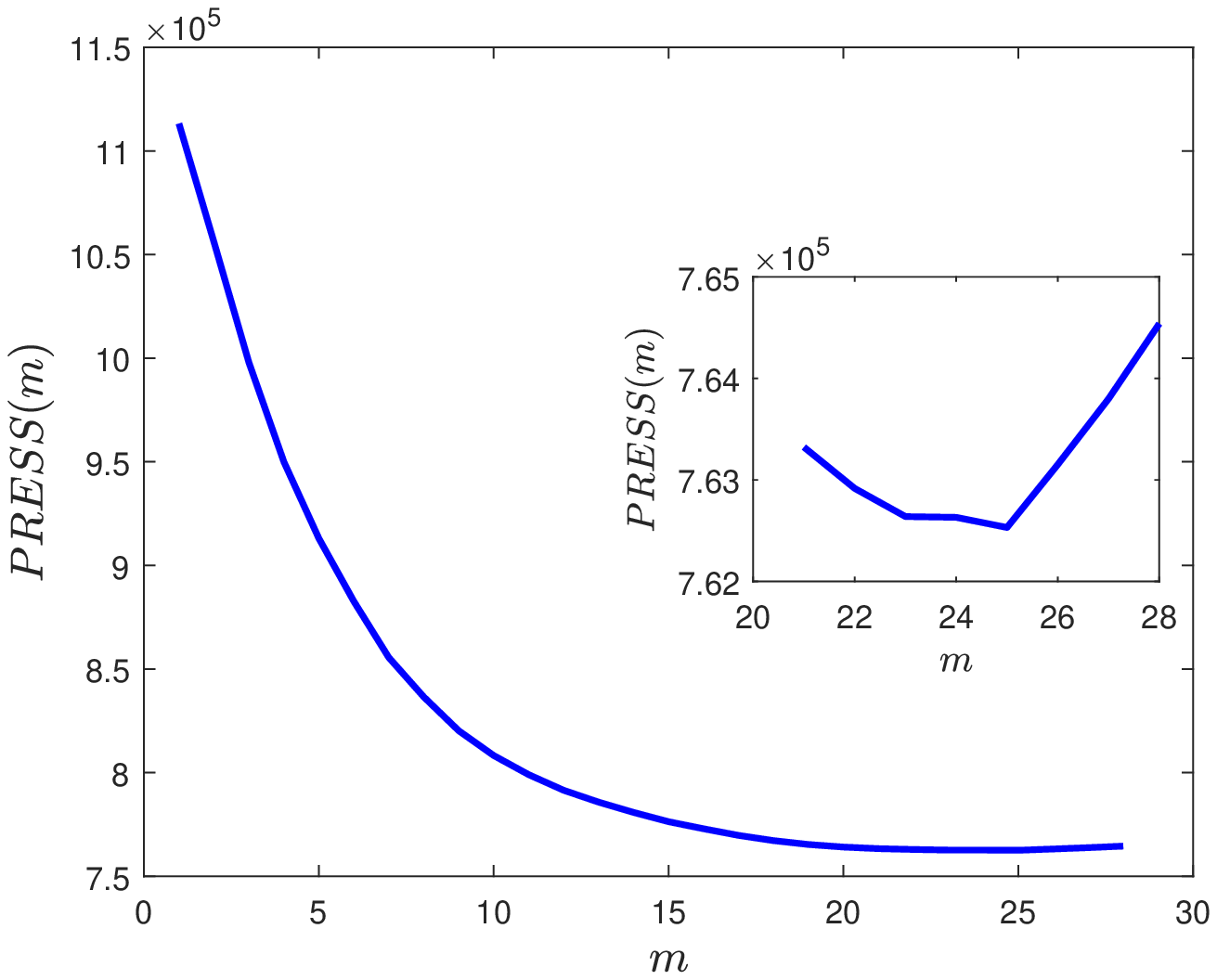}
\caption{\label{fig:crossval_simulation_het}heterogeneous}
\end{subfigure}
\caption{(Top) The median value of $\Lambda(m)$ (see Eq. \eqref{CumVarEqn}) for the cumulative variance method for the synthetic homogeneous and heterogeneous systems respectively. The $70\%$ and $90\%$ cutoff levels are indicated in dashed red lines, and occur at $m=12,22$  for the homogeneous system and $m=7,17$ for the heterogeneous one. (Bottom) For the homogenous and heterogeneous systems again, we plot the median of $\text{PRESS}(m)$ (see Eq. \eqref{PRESS_eqn}) using $10$ fold cross-validation for each sample. We see from the zoomed inset figures that the minimum $\text{PRESS}(m)$ occurs at $m=29$ for the homogenous system and $m=25$ for the heterogeneous one. }
\label{fig:mstar_comparison_literature}
\end{figure}

We compare our memory-based method, the cumulative variance method with $70\%$ and $90\%$ cutoffs and the $10$-fold cross-validation method for $100$ samples of the synthetic system described in Section \ref{Synthetic} and for the empirical dataset described in \ref{DataCleaning}, where the numerical outputs of $m^{\star}$ for these methods is detailed in the columns of Table \ref{tab:mstar_comparison}. 

\begin{table}[htbp]
  \centering
    \begin{tabular}{|c|c|c|c|}
    \hline
          & \multicolumn{2}{c|}{synthetic} & \multicolumn{1}{c|}{empirical} \\
    \hline
          & \multicolumn{1}{c|}{homogeneous} & \multicolumn{1}{c|}{heterogeneous} &  \\
    \hline
    {\bf memory-based} & {\bf 19}    & {\bf 12}    & {\bf 15} \\
    \hline
    cumulative variance & 12--22 & 7--17 & 13--27 \\
    \hline
    cross-validation & 29    & 25    & 28 \\
    \hline
          &       &       &  \\
    \hline
    $m_{max}$ & 30    & 28    & 35 \\
    \hline
    \end{tabular}%
  \caption{This table summarises the $m^{\star}$ values obtained for the synthetic data described in Section \ref{Synthetic} and empirical dataset described in \ref{DataCleaning}. Results from our memory-based method from Section \ref{Synthetic_results} are included in the first row. In the second row, we have the cumulative variance rule for the cutoffs $70\%$ and $90\%$. The final row includes the $\text{PRESS}(m)$ (see Eq. \eqref{PRESS_eqn}), using $10$-fold cross-validation.}
  \label{tab:mstar_comparison}%
\end{table}%

\begin{table}[htbp]
  \centering
    \begin{tabular}{|c|c|c|c|}
    \hline
          & \multicolumn{2}{c|}{synthetic} & \multicolumn{1}{c|}{empirical} \\
    \hline
          & \multicolumn{1}{c|}{homogeneous} & \multicolumn{1}{c|}{heterogeneous} &  \\
    \hline
    {\bf memory-based} & {\bf 138.6} & {\bf 137.6} & {\bf 209.7} \\
    \hline
    cross-validation & 1136.8 & 1146.3 & 1462.3 \\
    \hline
    \end{tabular}%
		\caption{Computational times in seconds for our proposed memory-based method (first row) and cross-validation using $10$ contiguous blocks (second row). The first two columns refer to the homogeneous and heterogeneous synthetic systems in Section \ref{Synthetic}. The final column is for the empirical dataset described in \ref{DataCleaning}. These performance times were calculated on a Windows $10$, CPU Intel i7-6700 3.4 GHz, RAM 16GB PC using MATLAB 2017b.}
  \label{tab:CompTime}%
\end{table}%

In Fig. \ref{fig:mstar_comparison_literature} (top panel), we plot for the homogeneous and heterogeneous synthetic data the median of $\Lambda(m)$ (see Eq. \eqref{CumVarEqn}) over all samples, indicating the $70\%$ and $90\%$ cutoffs in dashed red lines. The $70\%$ and $90\%$ cutoffs yield an optimal number of $12$ and $22$ components for the homogeneous case and $7$ and $17$ for the heterogeneous case, respectively. It makes sense that fewer components are needed in the heterogeneous case as more of the total variance is accounted for by the first principal components, which correspond by construction to the larger clusters. We recall that our memory-based method predicts $m^{\star}=19$ and $m^{\star}=12$ for the homogeneous and heterogeneous cases respectively, and these values fall squarely between the prescribed $70\%$ and $90\%$ cutoffs \cite{Jolliffe2002}. However, our method is superior in that it gives a unique value for $m^\star$ and not a range of values, and does not use subjective criteria or rules of thumb.

Fig. \ref{fig:mstar_comparison_literature} (bottom panel) depicts the median of $\text{PRESS}(m)$ across all samples, from which we see that the minimum occurs at $m^{\star}=29$ for the homogenous system and $m^{\star}=25$ for the heterogenous one. Hence, the cross-validation method would induce us to keep the majority of components in both systems. This is to be expected since cross-validation is based on minimising the out-of-sample prediction error (see Eq. \eqref{FactorModel}), hence performing the linear regression many times necessarily leads to a higher likelihood of including a larger number of principal components. This comes of course at the price of computational speed. Another interesting observation is that the minima occurring in both systems are not sharply defined, which indicates that the out-of-sample error made by including a larger number of components than the optimum $m^\star$ does not actually increase by a significant amount. 

Compared to cross-validation, our methodology leads to keeping fewer components. Our procedure, however, is less computationally expensive since it performs far fewer regressions to find $m^{\star}$ (see Table \ref{tab:CompTime}). Another advantage of our method over cross-validation can be spotted in the top panel of Fig. \ref{fig:mstar_comparison}, which highlights that only $9\%$ and $6\%$ of the total memory for the homogenous and heterogeneous systems after removing the market mode is unaccounted for to the right of $\hat\theta$. From the perspective of explaining the memory in the time series, therefore, our method does on average a very good job while requiring very limited computational resources.   

\begin{figure}
\begin{subfigure}{0.475\textwidth}
\centering
\includegraphics[width=\textwidth]{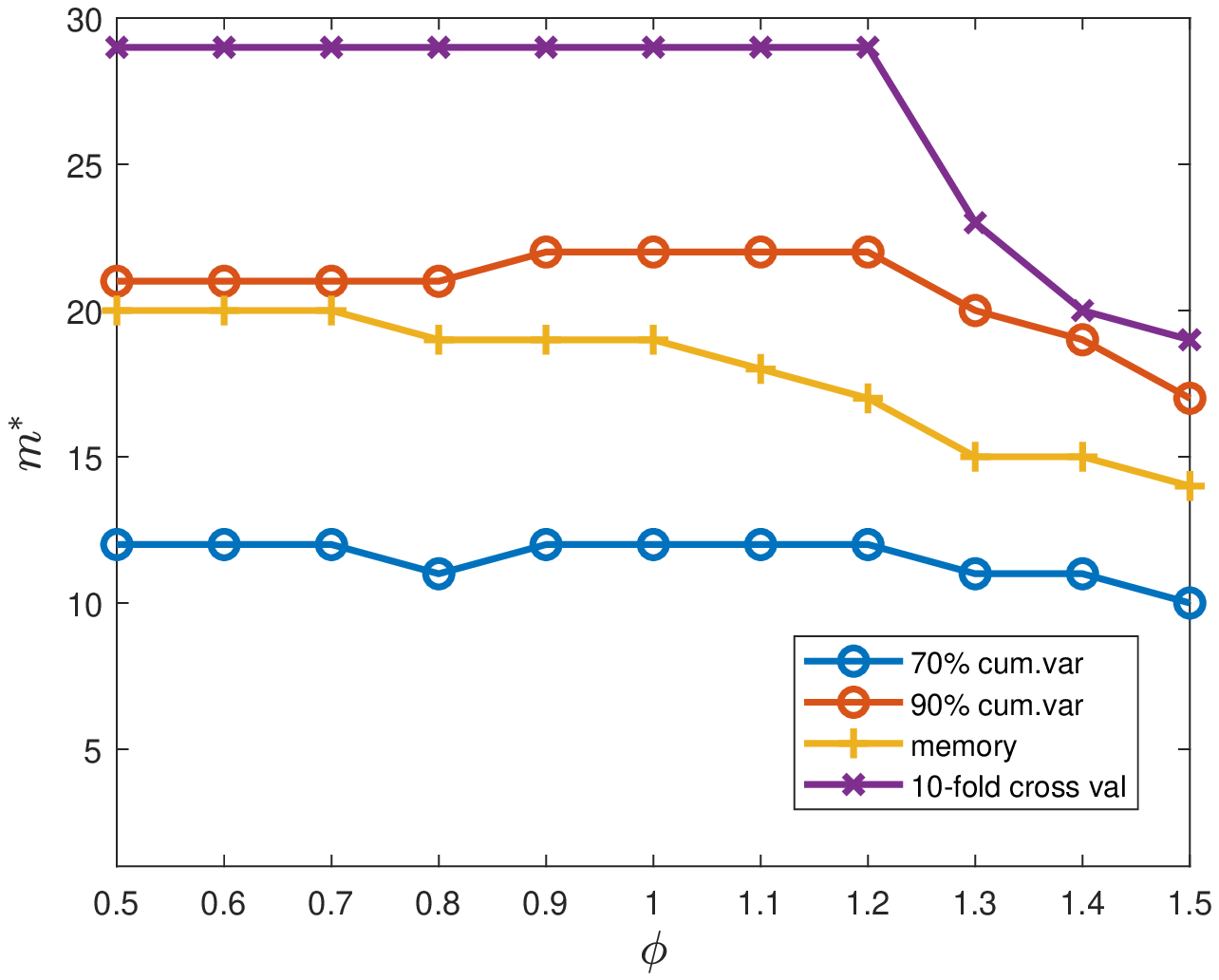}
\caption{\label{fig:optimalm_comparison_10fold}}
\end{subfigure}
\begin{subfigure}{0.475\textwidth}
\includegraphics[width=\textwidth]{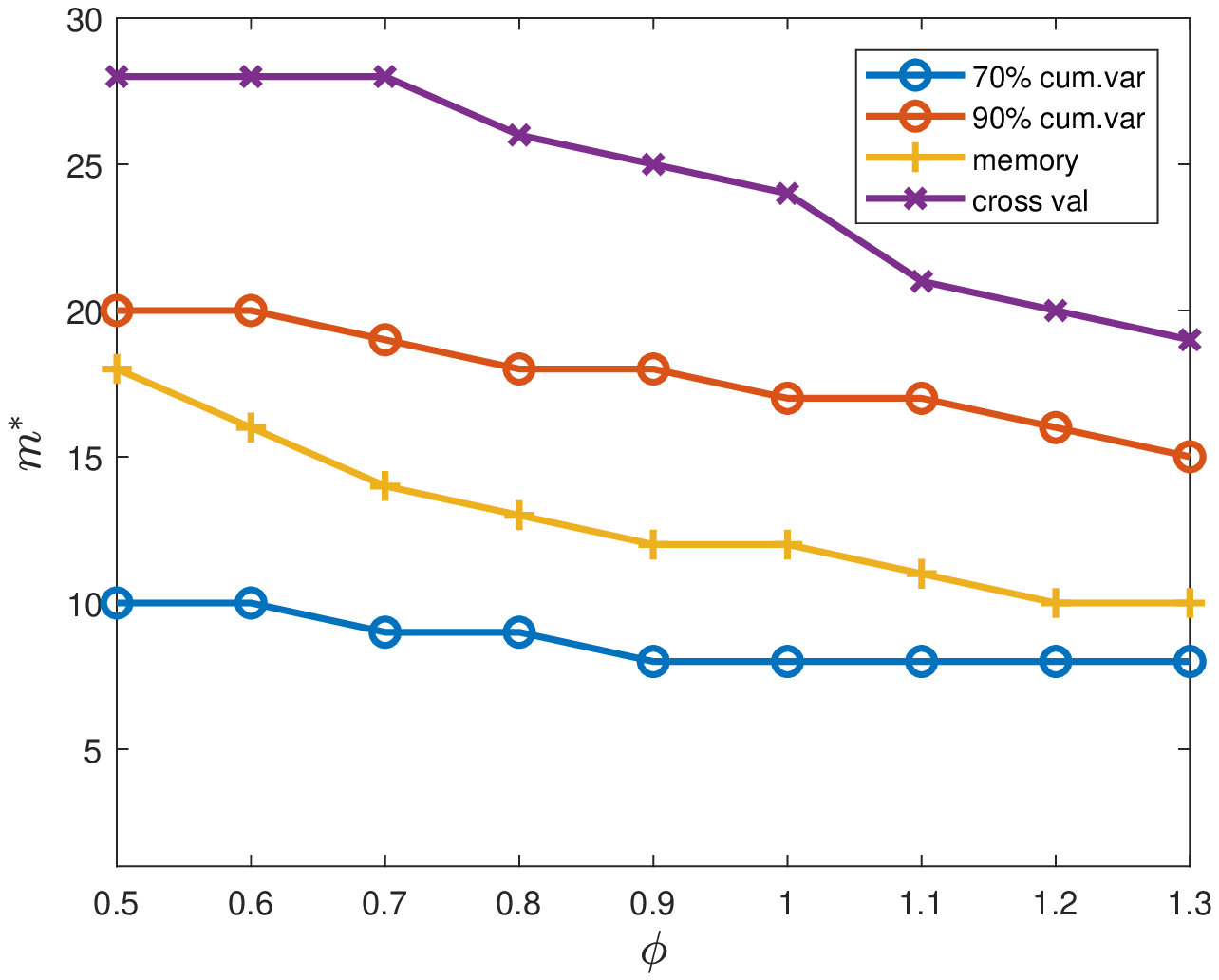}
\caption{\label{fig:optimalm_comparison_het}}
\end{subfigure}
\caption{A comparison of the different methods for selecting $m^{\star}$ by varying $\phi$, the noise level in the simulation of synthetic data (see Eq. \eqref{volproc}). For each value of $\phi$, $100$ samples of the process are generated, with the results for the homogeneous system plotted on the left, and for the heterogeneous system on the right. The blue and red lines represent the results for the $70\%$ and $90\%$ cumulative variance procedure. The orange line corresponds to our method. Finally the purple line represents results from $10$-fold cross-validation.}
\label{fig:optimalm_comparison}
\end{figure}
Now that we have compared the methods for a fixed $\phi$, the variance of the noise term for our synthetic data (see Eq. \eqref{volproc}), we can check how robust each of the methods is to changes in $\phi$. We note that fixing $\phi=1$ constitutes already a hard regime to analyse since it implies that the fluctuations due to $I_{k}(t)$ are of the same magnitude as the white noise, so we can see already that our method stands well compared to others with this high value of $\phi$. In Fig. \ref{fig:optimalm_comparison}, we compare -- using $100$ samples of the synthetic systems for the homogeneous and the heterogeneous cases -- the optimal value $m^\star$ predicted by the cumulative variance method with $70\%$ and $90\%$ cutoffs, the $10$-folds cross-validation method and our own memory-based procedure as we vary $\phi$. 

The $70\%$ and $90\%$ cutoffs for the cumulative variance rule remain relatively stable for most values of $\phi$, before slowly decreasing for higher values of $\phi$. This decrease occurs because the increased level of noise lowers the contribution to the variance from higher components, with the consequence that the cutoff is reached sooner for higher values of $\phi$. Within our memory-based method, the value of $m^\star$ decreases for increasing $\phi$. This decrease occurs because a higher amount of white noise increasingly masks the long-memory properties of the underlying signal, and will affect the deeper principal components more since they have a lower memory strength (lower $H_{k}$) anyway. This is a desirable property since it means that lowering the noise level will lead us to retain more principal components. Whilst the decrease in the number of components occurs earlier than for the cumulative variance method, it still remains between the $70\%$ and $90\%$ cutoffs, and even closer to the $90\%$ cutoff for lower values of $\phi$.

\begin{figure}
\begin{subfigure}{0.475\textwidth}
\centering
\includegraphics[width=\textwidth]{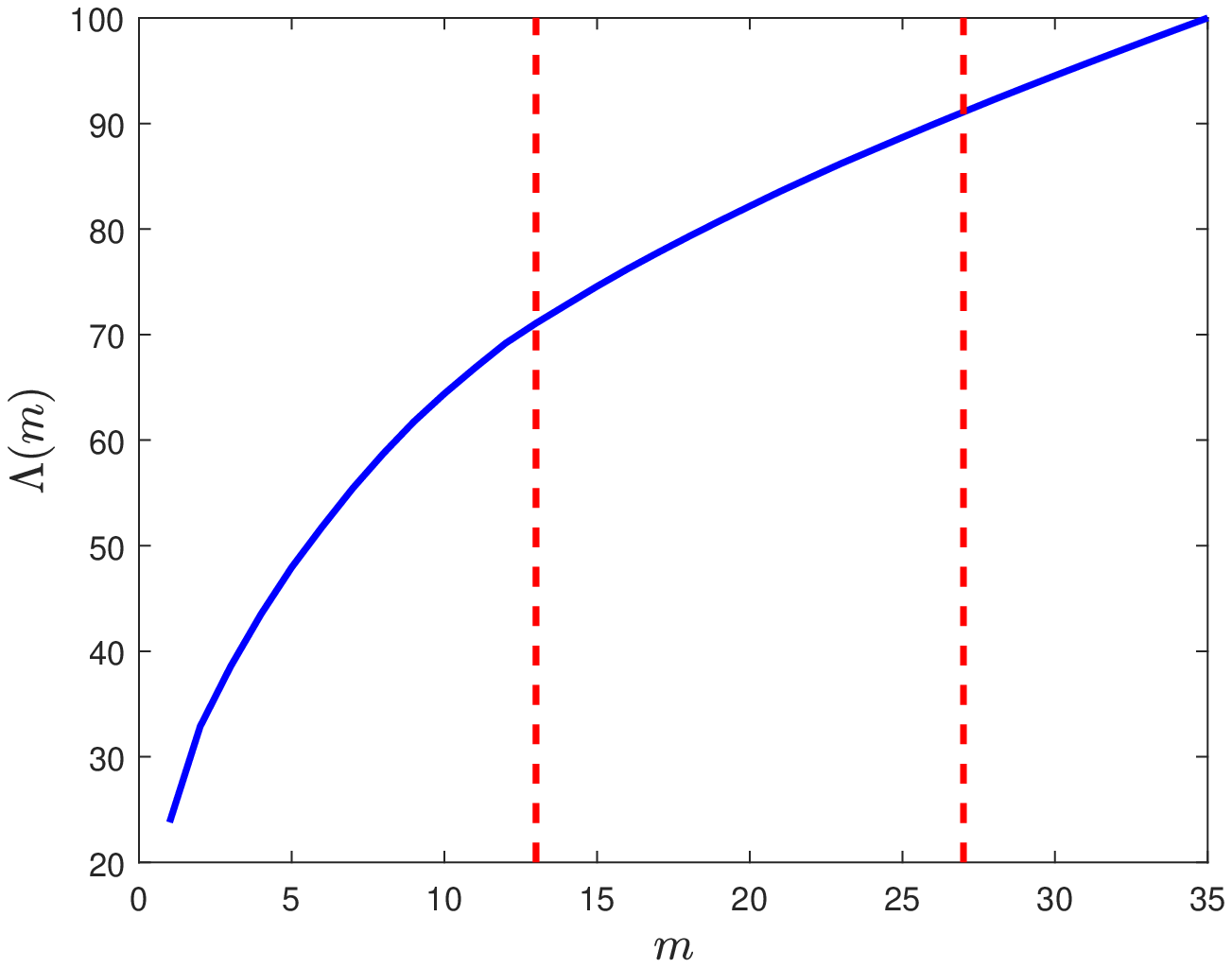}
\caption{\label{fig:CumExplainedVariance}}
\end{subfigure}
\begin{subfigure}{0.475\textwidth}
\centering
\includegraphics[width=\textwidth]{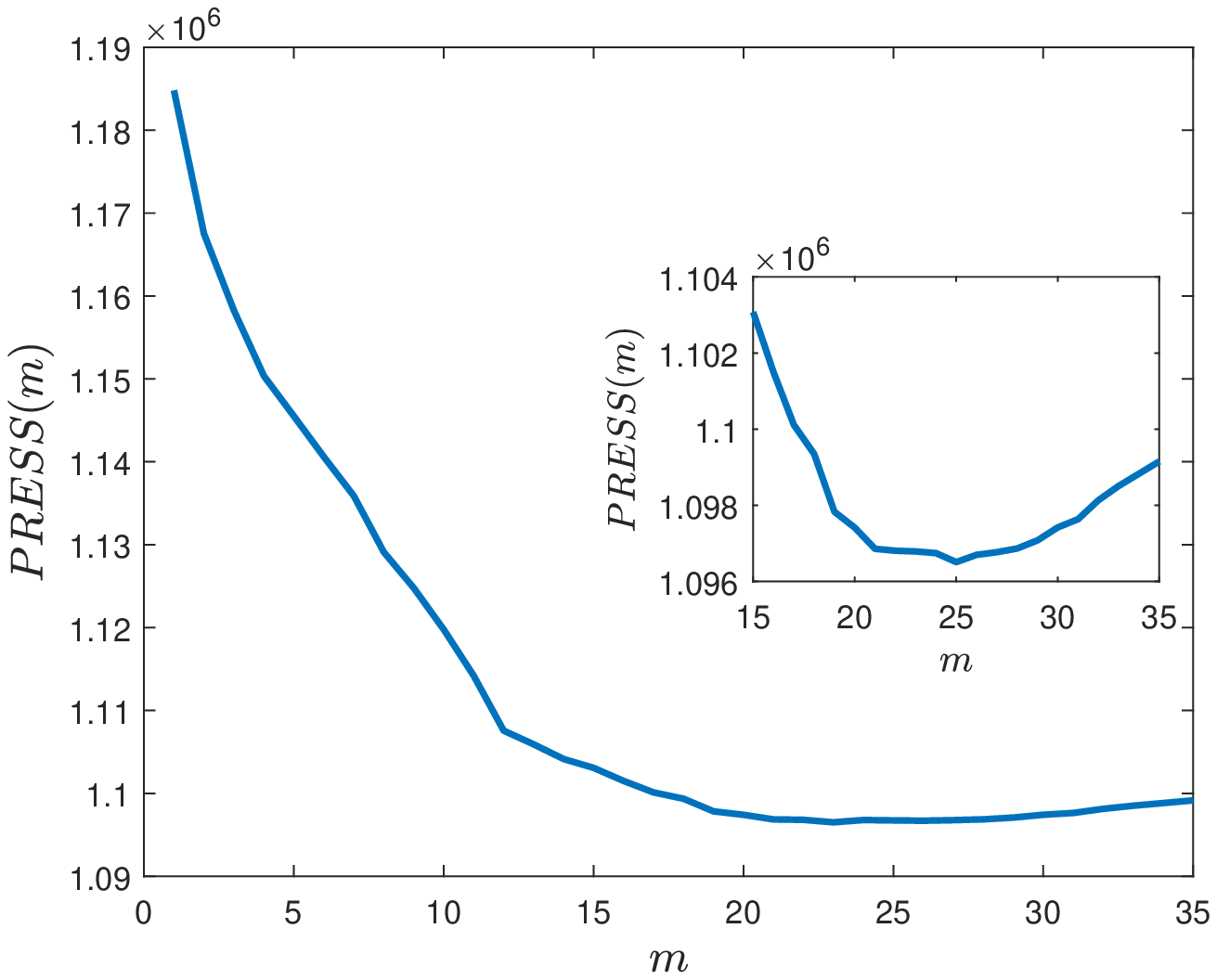}
\caption{\label{fig:CrossVal}}
\end{subfigure}
\caption{Comparison between the cumulative variance rule, cross-validation and our memory-based method of determining $m^{\star}$ applied to the empirical dataset defined in \ref{DataCleaning}. (Left) Plot of $\Lambda(m)$ defined in Eq. \eqref{CumVarEqn} with the red dashed lines at $m=13$ and $m=27$ indicating the region where between $70\%$ and $90\%$ of the total variance is explained by the principal components. (Right) Plot of $\text{PRESS}(m)$ given in Eq. \eqref{PRESS_eqn} using $10$-fold cross-validation, with a zoomed in inset version showing that the minimum occurs at $m=28$.}
\label{fig:comparison_methods_empirical}
\end{figure}

For the empirical dataset, described in \ref{DataCleaning}, we plot in Fig. \ref{fig:comparison_methods_empirical} (left) the plot of $\Lambda(m)$, the cumulative percentage of variation explained by the $m$ principal components. We see that if we set our target between $70\%$ and $90\%$ of the cumulative variance as prescribed in \cite{Jolliffe2002}, this will correspond to retaining between $13$ and $27$ components, but again it is not clear a priori what exact value within this range we should pick. In Fig. \ref{fig:comparison_methods_empirical} (right), we plot $\text{PRESS}(m)$ obtained via $10$-fold cross-validation, in which the minimum occurs at $m^{\star}=28$, close to the $90\%$ cutoff for the cumulative variance. Again -- compared to cross-validation -- our method picks out fewer principal components, but we obtain our result in far less computational time (see Table \ref{tab:CompTime}), and with $m^{\star}=15$ we can already account for $80\%$ of the memory.    

\section{Conclusion} \label{Conclusion}
In this paper, we have proposed a novel, data-driven method to select the optimal number $m^\star$ of principal components to retain in the Principal Component Analysis of data with long memory. The main steps are detailed in Section \ref{Method}. We used the crucial fact that subsequent components contribute a decreasing amount to the total memory of the system. This allows us to identify a unique, non-subjective and computationally inexpensive stopping criterion, which compares very well with other available heuristic procedures such as cumulative variance and cross-validation (see Tables \ref{tab:mstar_comparison} and \ref{tab:CompTime}). We tested our method on two synthetic systems: a homogeneous and heterogeneous version \ref{Synthetic}, and also on an empirical dataset of financial log-volatilities, described in \ref{DataCleaning}. Our results could be applied to any large dataset endowed with long-memory properties, for example in climate science \cite{Von2001,Franzke2012} and neuroscience \cite{Linkenkaer2001,Pang2016}. A potential direction for future work could be using a null hypothesis for the bulk eigenvalues which takes into account the presence of autocorrelations rather than the MP distribution used here. A comparison with the cluster driven method presented in \cite{Verma2017} or extending the method for example to nonlinear PCA \cite{Karhunen1994} could also be explored.

\appendix

\section{\\Empirical Dataset} \label{DataCleaning}
The empirical dataset we shall use consists of the daily closing prices of 1270 stocks in the New York Stock Exchange (NYSE), National Association of Securities Dealers Automated Quotations (NASDAQ) and American Stock Exchange (AMEX) from $1$st January 2000 to $12$th May 2017, which amounts to $4635$ entries for each price time series. We make sure that the stocks are ``aligned" through the data cleaning procedure described here. A typical source of misalignment is the fact that some stocks have not been traded on certain days. To ensure we keep as many entries as possible, we fill the gaps dragging the last available price ahead and assuming that a gap in the price time-series corresponds to a zero log-return. At the same time, we do not wish to drag ahead too many prices as doing so would compromise the statistical significance of the time-series. The detailed procedure goes as follows:
\begin{enumerate}
\item Remove from the dataset the price time-series with length smaller than $p$ times the longest one;
\item Find the common earliest day among the remaining time-series;
\item Create a reference time-series of dates when at least one of the stocks has been traded starting from the earliest common date found in the previous step;
\item Compare the reference time-series of dates with the time-series of dates of each stock and fill the gaps dragging ahead the last available price.
\end{enumerate}
In this paper, we chose $p=0.90$ to ensure that we keep the time-series as unmodified as possible. For example, the common earliest day for our dataset is $3$rd of January 2000. In this period, the stock Ameris Bancorp (ABCB), was not traded on $35$ days in the time period and therefore the last available price was used to fill these particular days. Another example is the stock Allied Healthcare Products (AHPI), which was not traded for $508$ days in the time period we study, and is removed since its length is less than $p$ times the longest time series. However, the results do not change if we pick a higher value of $p$. Applying this procedure leaves our dataset with $N=1202$ stocks. Hence $\bm{X}$ and $\bm{X}^{(\text{market})}$ are $4364\times 1202$ matrices.

\section{\\Financial interpretation of the eigenvectors and portfolio optimisation} \label{PortfolioOpt}
Another motivation for the application of PCA to financial correlation matrices is the financial interpretation of the first principal components, which we explain here. First, we recall that the empirical correlation matrix $\bm{E}$ between the standardised log volatilities is defined as
\begin{equation}
E_{ij}=\frac{1}{T}\sum_{t=1}^{T}\ln |r_{i}(t)|\ln |r_{j}(t)| \ .
\end{equation} 
We call $\bm{w}_{m}$ the eigenvectors of $\bm{E}$ with $\lambda_m$ its associated eigenvalue. We interpret the entries of $\bm{w}_{m}$ as the weights of a portfolio, with $w_{im}>0$ indicating a long position where we buy the stock in the expectation that its value will rise, and $w_{im}<0$ denoting a short position where we expect the stock's value to fall and hence sell it \cite{Bouchaud2009}.

The covariance between the log volatilities of the portfolios $m$ and $m'$ is:
\begin{equation}
\frac{1}{T}\sum_{t=1}^{T}\left(\sum_{i=1}^{N}w_{im}\ln |r_{i}(t)|\right)\left(\sum_{j=1}^{N}w_{jm'}\ln |r_{j}(t)|\right)=\sum_{m'}\lambda_{m}\delta_{m,m'} \ , \label{PortfolioCovariance}
\end{equation}
where $w_{im}$ and $w_{jm'}$ are the entries of the eigenvector $\bm{w}_{m}$ and $\bm{w}_{m'}$ respectively. Hence the returns defined by the portfolio $\bm{w}_{m}$ and another eigenvector $\bm{w}_{m'}$ are uncorrelated. Another consequence of Eq. \ref{PortfolioCovariance} is that the variance of the returns, which is used to measure the risk of a portfolio, is the eigenvalue $\lambda_{m}$. Hence larger eigenvalues of the portfolio defined by $\bm{w}_{m}$ have a higher risk. Knowing this information about the eigenvalues and their corresponding eigenvectors can therefore inform an investment manager in deciding how to pick portfolios both individually and to reduce a set of portfolios' overall risk by using orthogonal portfolios defined by $\bm{w}_{m}$.

For a given level $\Delta$ of tolerable risk, we can also find the optimal investment weights $\bm{w}_{\text{opt}}$ by solving the minimisation problem
\begin{align}
&\min_{\bm{w}}\bm{w}^{T}\bm{E}\bm{w} \\
&\text{such that   } \bm{X}\bm{w}=\Delta \ .
\end{align}
This is known as Markowitz portfolio optimisation theory \cite{Markowitz1952}, and can be solved via Lagrange multipliers to give
\begin{equation}
\bm{w}_{\text{opt}}=\Delta\frac{\bm{E}^{-1}\bm{R}}{\bm{R}^{\dagger}\bm{E}^{-1}\bm{R}} \ , \label{MarkowitzSoln}
\end{equation} 
 with $\bm{w}_{\text{opt}}$ indicating the optimal portfolio weight. We see that the distribution of the eigenvalues enters the portfolio optimisation through the inverse matrix $\bm{E}^{-1}$ in Eq. \eqref{MarkowitzSoln}. Normally, Eq. \eqref{MarkowitzSoln} is applied directly by simply using the sample estimator $\bm{E}$. However, since $\bm{E}$ is empirical, it is subject to noise inherent in the data which means it is vulnerable to the noisy distribution of the eigenvalues, in turn causing the $\bm{w}_{opt}$ found to underestimate risk \cite{Bun2017}.

\begin{figure}
\centering
\includegraphics[width=0.75\textwidth]{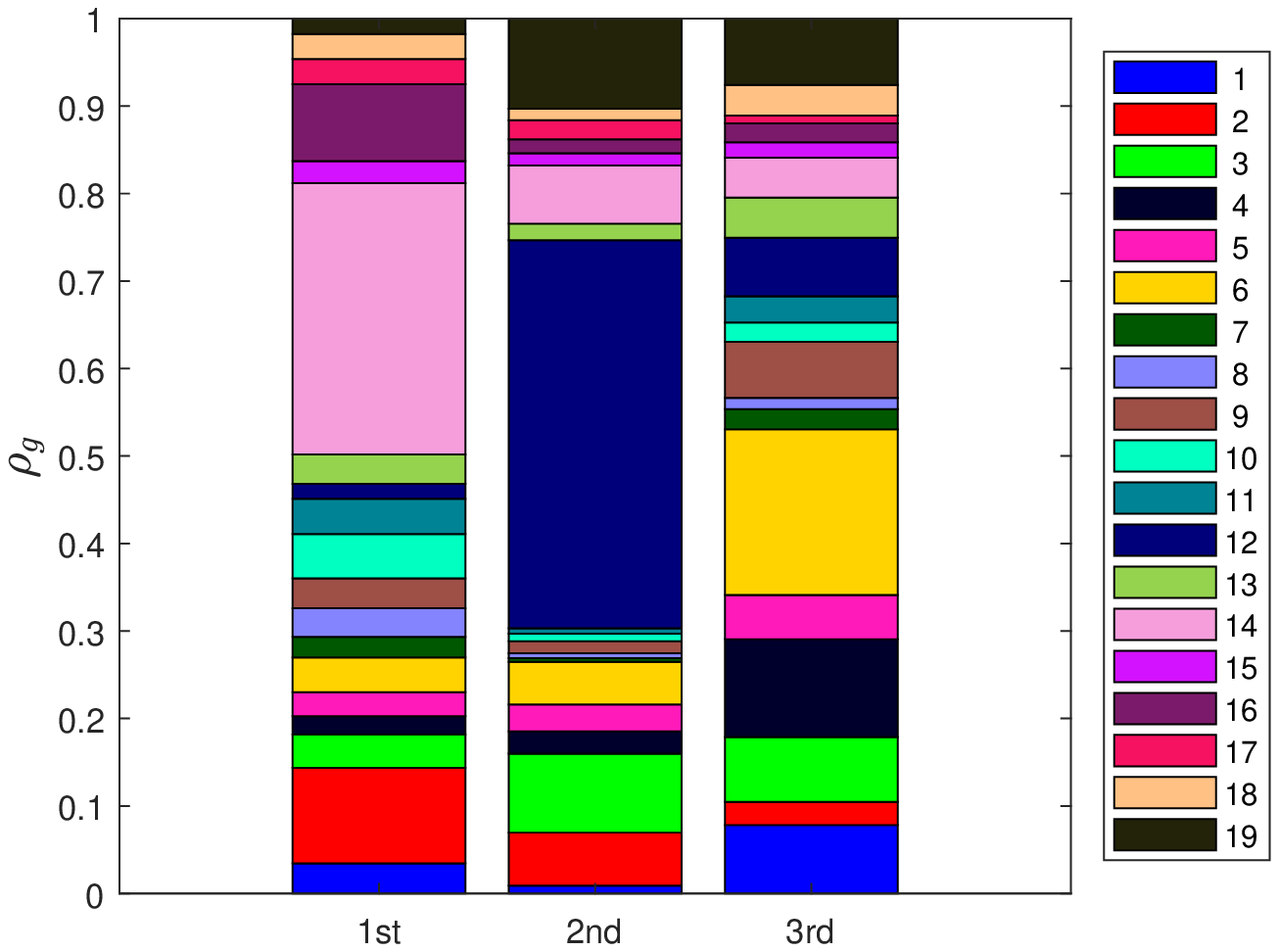}
\caption{Plots the $\rho_{g}$ defined in \ref{PortfolioOpt}, which is the projection of the eigenvector onto the ICB supersector groups, for the eigenvectors of the first three principal components of $\bm{G}$ for the data detailed in \ref{DataCleaning}. The legend corresponds to each of the ICB supersector groups.}
\label{fig:Composition_evectors}
\end{figure}

We also note that in line with \cite{Plerou2002}, the eigenvectors corresponding to these eigenvalues beyond the MP bulk for $\bm{G}$ for the empirical data in \ref{DataCleaning} can be identified as belonging to particular or a mixture of $19$ economic Industrial Classification Benchmark (ICB) supersectors \cite{ICB}. We can quantify this for the eigenvectors of $\bm{G}$ given in Eq. \eqref{GEqn}, $\bm{v}_{i}$, by defining a $19$-dimensional vector $\rho_{i}$, with entries $\rho_{g,i}$, $g=1,...,19$. Specifically, we define a projection matrix $\bm{P}$ with entries
\[P_{ig} = \begin{cases}
						1/N_{g} & \text{if $i$ is in supersector $g$} \\
						0       & \text{else}\ ,	
						\end{cases}
\]
where $N_{g}$ is the number of stocks that are part of supersector $g$. From this we can define $\rho_{i}$ as
\begin{equation}
\rho_{i}=\gamma_{i}\bm{P}\bm{v}_{i} \ ,
\end{equation}
where $\gamma_{i}$ is the normalisation constant $\sum_{g=1}^{19}\rho_{g,i}$. Each $\rho_{g,i}$ gives the contribution of the $g$-th ICB supersector to the $ith$ eigenvector. We plot $\rho_{g}$ for the first three eigenvectors in Fig. \ref{fig:Composition_evectors}. We can see that each eigenvector is dominated by the Real Estate (colour 14), Oil and Gas (colour 1) and Financial Services (colour 6) respectively for the first, second and third principal components. 

\section{\\Lasso regression} \label{Lasso}
Lasso regression is used to find the values of the coefficients $\beta_{ip}$ using Eq. \eqref{FactorModel}. Further details of the use of this method is provided in this appendix. Lasso regression \cite{Tibshirani1996} provides a way of dealing with overfitting explanatory variables (in our case $I_{k}(t)$) and also of performing feature selection, which takes into account a stock $i$'s log-volatility not being affected by changes of $I_{k}(t)$ . Lasso regression solves the constrained minimisation problem
\begin{equation}
\min_{\boldsymbol{\beta}_{i}} \frac{1}{T}\sum_{t=1}^{T}\left(c_{i}(t)-\bm{I}(t)^{\dagger}\boldsymbol{\beta}_{i}\right)^{2}+\Upsilon P_{a}(\boldsymbol{\beta}_{i}) \ ,
\end{equation}
where $\boldsymbol{\beta}_{i}$ is the vector of loadings given by $(\beta_{i1}, \beta_{i2}, \dots,\beta_{im_{\text{max}}})^{\dagger}$, $\bm{I}(t)$ is the matrix whose columns are $(I_{1}(t),I_{2}(t), \dots, I_{m_{\text{max}}}(t))$ and $\Upsilon$ is a hyperparameter. $P_{a}(\boldsymbol{\beta}_{i})$ is defined as
\begin{equation}
P_{a}(\boldsymbol{\beta}_{i})=\sum_{j=1}^{m_{\text{max}}}|\beta_{ij}| \ . \label{LassoPenalty}
\end{equation}  
The sum in Eq. \eqref{LassoPenalty} is the $\mathcal{L}_{1}$ penalty for the lasso regression. The $\Upsilon$ controls the amount of regularisation: the higher it is, the more loadings are zero. To find $\Upsilon$, we set its scale according to \cite{Friedman2010} and use 10 cross-validated fits \cite{Tibshirani1996}, picking the $\Upsilon$ that gives the minimum prediction error. We have also investigated the stability of the results with respect to changes in $\Upsilon$, and altering the penalty in \eqref{LassoPenalty} to a L2 penalty. In either case there is little difference to the calculated $m^{\star}$ values.

\section{\\Fitting Procedure for $\theta$} \label{ThetaHatFitting}
We can estimate $\theta$ by assessing what region of $\zeta(m)$ is most linear in log-log scale, which is done by assessing on each interval $m=\tilde{\theta},...,m_{\mathrm{max}}$, where $\tilde{\theta}=2,...,m_{\mathrm{max}}$, the quality of a linear fit in log-log scale of $\zeta(m)$ in this interval. The estimate of $\theta$, $\hat{\theta}$ is then the value of $\tilde{\theta}$ that gives the best-quality linear fit. To assess the quality of the fit, we use the adjusted $R_{\mathrm{adj}}^{2}$ value \cite{Theil1958}:
\begin{equation}
R_{\mathrm{adj}}^{2}=1-(1-R^{2})\frac{n-1}{n-2} \ , \label{AdjR}
\end{equation}
where $R^{2}$ is the normal coefficient of determination \cite{Draper2014}, and $n$ is the size of the interval. Note we have written the formula for our specific case where the number of explanatory variables is $1$. If $R_{\mathrm{adj}}^{2}$ is higher, then the interval $m=\tilde{\theta},...,m_{\mathrm{max}}$ is better described by a linear trend. The difference between $R_{\mathrm{adj}}^{2}$ and $R^{2}$ is that the former can take into account the different sample sizes induced by the differently sized intervals by reducing the value obtained through $R^{2}$ for smaller values of $n$. $\hat{\theta}$ is then given by 
\begin{equation}
\hat{\theta}=\max_{\tilde{\theta}}R_{\mathrm{adj}}^{2}(\tilde{\theta}) \ , \label{ThetaHat}
\end{equation}
which is the value of $\tilde{\theta}$ which maximises $R_{adj}^{2}$ and gives the region of best-quality linear fit.

\section{\\ Exponentially decaying autocorrelations} \label{Autoregressive}
\begin{figure}
\begin{subfigure}{0.475\textwidth}
\centering
\includegraphics[width=\textwidth]{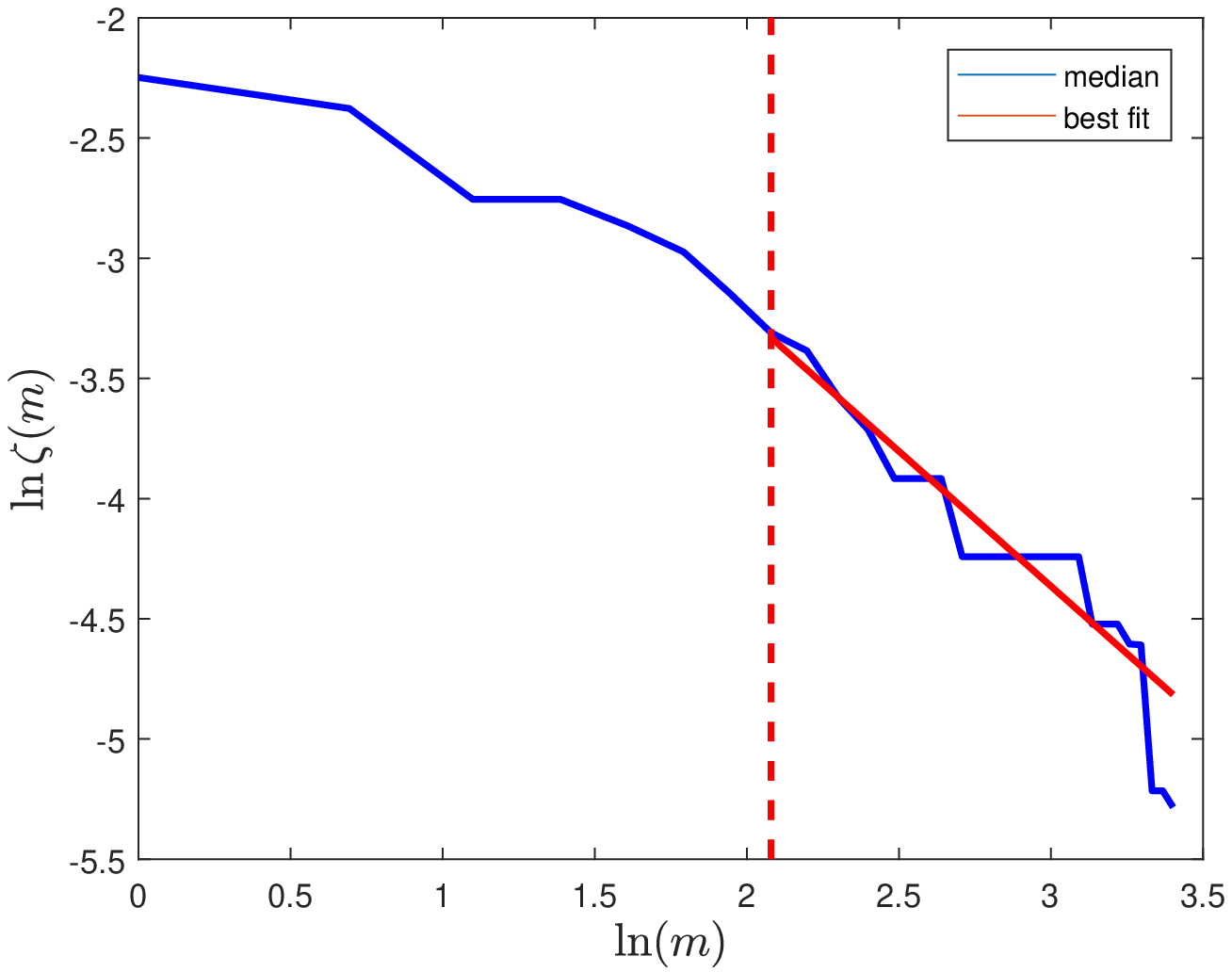}
\caption{\label{fig:loglog_samples_medianratio_AR} homogeneous}
\end{subfigure}
\begin{subfigure}{0.475\textwidth}
\centering
\includegraphics[width=\textwidth]{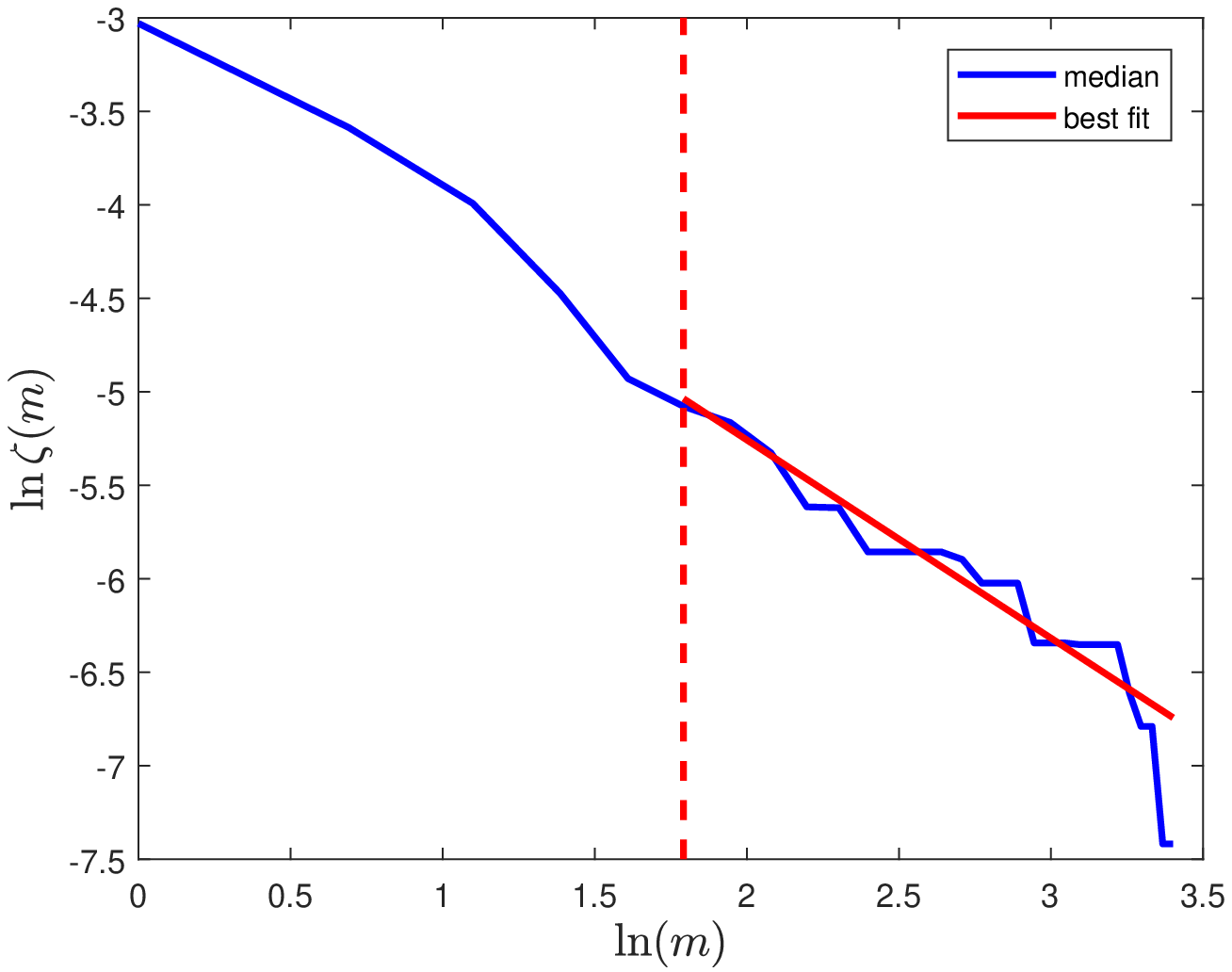}
\caption{\label{fig:loglog_samples_medianratio_het_AR} heterogeneous}
\end{subfigure}
\caption{(Top left) Plot of $\ln(\zeta(m))$ vs $\ln(m)$ for the same homogeneous synthetic system described in the original manuscript but using AR(1) as the generating process for $I_{k(i)}(t)$. The blue line is the value of $\zeta(m)$ across all assets, with the dashed red line indicating $\hat{\theta}=8$, the point at which the concavity changes. (Top right) Same plot but for the same heterogeneous simulated system described in the original manuscript, where $\hat{\theta}=6$.}
\label{fig:mstar_comparison_AR}
\end{figure}

The Autoregressive process of order $1$ (AR(1)) is given by \cite{Box2015}
\begin{equation}
X(t)=\epsilon(t)+ \psi X(t-1) \ , \label{AREqn}
\end{equation}
where $\epsilon(t),\epsilon(t-1),...$ are all white noise terms, $\psi$ is the autoregressive parameter. To enforce stationarity and positive autocorrelations note that we must have that $0<\psi<1$ \cite{Box2015}. The presence of the second term in Eq. \eqref{AREqn} introduces memory into the process. The autocorrelation function of $X(t)$ is known to be exponential \cite{Box2015}, with increasing $\psi$ increasing the strength of the memory, in contrast to the FBM we used in section \ref{Synthetic}. By using AR(1) to generate $I_{0}$ and the set of $I_{k(i)}(t)$ with parameters $\psi_{0}$ and $\psi_{k}$ respectively, we can investigate whether the method proposed here is still valid when the autocorrelation decays exponentially. For $I_{0}(t)$, we fix $\psi_{0}=0.95$. Each $I_{k(i)}(t)$ is generated using an equally spaced vector from $0.65$ to $0.95$ for $\psi_{k}$ set in a similar way described in section \ref{Synthetic} to reflect the empirical result of \cite{Micciche2013}. We then repeat the steps given in section \ref{ProcedureSummary} for the same homogenous and heterogenous synthetic systems described in section \ref{Synthetic}. The log-log plots of $\zeta(m)$ vs $m$ are detailed in Fig. \ref{fig:mstar_comparison_AR}. We see that for both systems whilst we do see a decrease, it is not accurately described by a straight line in log-log scales in this case, as compared to Figs. \ref{fig:loglog_samples_medianratio} and \ref{fig:loglog_samples_medianratio_het}. Therefore we can conclude that whilst our method can be applied also in the case of faster, exponentially decaying autocorrelation, it is less precise. 

\section*{Acknowledgements}
We thank Bloomberg for providing the data used in this paper. We also wish to thank the ESRC Network Plus project 'Rebuilding macroeconomics'. We acknowledge support from Economic and Political Science Research Council (EPSRC) grant EP/P031730/1. We are grateful to the NVIDIA corporation for supporting our research in this area with the donation of a GPU.

\section*{\textemdash \textemdash \textemdash \textemdash \textemdash \textendash{}}

\bibliographystyle{unsrt}
\bibliography{PCA_paper}

\begin{thebibliography}{10}

\bibitem{Jolliffe2002}
Ian Jolliffe.
\newblock {\em Principal component analysis}.
\newblock Wiley Online Library, 2002.

\bibitem{Jackson1993}
Donald~A Jackson.
\newblock Stopping rules in principal components analysis: a comparison of
  heuristical and statistical approaches.
\newblock {\em Ecology}, 74(8):2204--2214, 1993.

\bibitem{Sonka2014}
Milan Sonka, Vaclav Hlavac, and Roger Boyle.
\newblock {\em Image processing, analysis, and machine vision}.
\newblock Cengage Learning, 2014.

\bibitem{Stein2006}
Sarah A~Mueller Stein, Anne~E Loccisano, Steven~M Firestine, and Jeffrey~D
  Evanseck.
\newblock Principal components analysis: a review of its application on
  molecular dynamics data.
\newblock {\em Annual Reports in Computational Chemistry}, 2:233--261, 2006.

\bibitem{Pang2016}
Rich Pang, Benjamin~J Lansdell, and Adrienne~L Fairhall.
\newblock Dimensionality reduction in neuroscience.
\newblock {\em Current Biology}, 26(14):R656--R660, 2016.

\bibitem{Alexander2002}
Carol Alexander.
\newblock Principal component models for generating large garch covariance
  matrices.
\newblock {\em Economic Notes}, 31(2):337--359, 2002.

\bibitem{Bun2017}
Jo{\"e}l Bun, Jean-Philippe Bouchaud, and Marc Potters.
\newblock Cleaning large correlation matrices: tools from random matrix theory.
\newblock {\em Physics Reports}, 666:1--109, 2017.

\bibitem{Darbyshire2016}
J.H.M. Darbyshire.
\newblock {\em The Pricing and Trading of Interest Rate Derivatives: A
  Practical Guide to Swaps}.
\newblock Aitch and Dee Limited, 2016.

\bibitem{Maaten2009}
Laurens Van Der~Maaten, Eric Postma, and Jaap Van~den Herik.
\newblock Dimensionality reduction: a comparative.
\newblock {\em J Mach Learn Res}, 10:66--71, 2009.

\bibitem{Aste2005}
Tomaso Aste, Tiziana Di~Matteo, and ST~Hyde.
\newblock Complex networks on hyperbolic surfaces.
\newblock {\em Physica A: Statistical Mechanics and its Applications},
  346(1-2):20--26, 2005.

\bibitem{Tumminello2005}
Michele Tumminello, Tomaso Aste, Tiziana Di~Matteo, and Rosario~N Mantegna.
\newblock A tool for filtering information in complex systems.
\newblock {\em Proceedings of the National Academy of Sciences},
  102(30):10421--10426, 2005.

\bibitem{Matteo2010}
Tiziana Di~Matteo, Francesca Pozzi, and Tomaso Aste.
\newblock The use of dynamical networks to detect the hierarchical organization
  of financial market sectors.
\newblock {\em The European Physical Journal B}, 73(1):3--11, 2010.

\bibitem{Aste2012}
Tomaso Aste, Ruggero Gramatica, and Tiziana Di~Matteo.
\newblock Exploring complex networks via topological embedding on surfaces.
\newblock {\em Physical Review E}, 86(3):036109, 2012.

\bibitem{Barfuss2016}
Wolfram Barfuss, Guido~Previde Massara, Tiziana Di~Matteo, and Tomaso Aste.
\newblock Parsimonious modeling with information filtering networks.
\newblock {\em Physical Review E}, 94(6):062306, 2016.

\bibitem{Verma2017}
Anshul Verma, Riccardo~Junior Buonocore, and Tiziana Di~Matteo.
\newblock A cluster driven log-volatility factor model: a deepening on the
  source of the volatility clustering.
\newblock {\em Quantitative Finance}, pages 1--16, 2018.

\bibitem{Hinton1994}
Geoffrey~E Hinton and Richard~S Zemel.
\newblock Autoencoders, minimum description length and helmholtz free energy.
\newblock In {\em Advances in neural information processing systems}, pages
  3--10, 1994.

\bibitem{Regier2016}
Jeffrey Regier and Jon McAuliffe.
\newblock Second-order stochastic variational inference.
\newblock In {\em Bay Area Machine Learning Symposium (BayLearn)}, 2016.

\bibitem{Comon1994}
Pierre Comon.
\newblock Independent component analysis, a new concept?
\newblock {\em Signal processing}, 36(3):287--314, 1994.

\bibitem{Hyvarinen2004}
Aapo Hyv{\"a}rinen, Juha Karhunen, and Erkki Oja.
\newblock {\em Independent component analysis}, volume~46.
\newblock John Wiley \& Sons, 2004.

\bibitem{Cattell1966}
Raymond~B Cattell.
\newblock The scree test for the number of factors.
\newblock {\em Multivariate behavioral research}, 1(2):245--276, 1966.

\bibitem{Sugiyama1976}
T~Sugiyama and Howell Tong.
\newblock On a statistic useful in dimensionality reduction in multivariable
  linear stochastic system.
\newblock {\em Communications in statistics-theory and methods}, 5(8):711--721,
  1976.

\bibitem{Huang1992}
Deng-Yuan Huang and Sheng-Tsaing Tseng.
\newblock A decision procedure for determining the number of components in
  principal component analysis.
\newblock {\em Journal of statistical planning and inference}, 30(1):63--71,
  1992.

\bibitem{Bartlett1950}
Maurice~S Bartlett.
\newblock Tests of significance in factor analysis.
\newblock {\em British Journal of Mathematical and Statistical Psychology},
  3(2):77--85, 1950.

\bibitem{Ali1985}
Asghar Ali, GM~Clarke, and K~Trustrum.
\newblock Principal component analysis applied to some data from fruit
  nutrition experiments.
\newblock {\em The Statistician}, pages 365--370, 1985.

\bibitem{Wold1978}
Svante Wold.
\newblock Cross-validatory estimation of the number of components in factor and
  principal components models.
\newblock {\em Technometrics}, 20(4):397--405, 1978.

\bibitem{Eastment1982}
HT~Eastment and WJ~Krzanowski.
\newblock Cross-validatory choice of the number of components from a principal
  component analysis.
\newblock {\em Technometrics}, 24(1):73--77, 1982.

\bibitem{Beran2017}
Jan Beran.
\newblock {\em Statistics for long-memory processes}.
\newblock Routledge, 2017.

\bibitem{Cont2001}
Rama Cont.
\newblock Empirical properties of asset returns: stylized facts and statistical
  issues.
\newblock {\em Quantitative Finance}, 1:223--236, 2001.

\bibitem{Hull1987}
John Hull and Alan White.
\newblock The pricing of options on assets with stochastic volatilities.
\newblock {\em The journal of finance}, 42(2):281--300, 1987.

\bibitem{Hull2006}
John~C Hull.
\newblock {\em Options, futures, and other derivatives}.
\newblock Pearson Education India, 2006.

\bibitem{Bouchaud2009b}
Jean-Philippe Bouchaud and Marc Potters.
\newblock {\em Theory of financial risk and derivative pricing: from
  statistical physics to risk management}.
\newblock Cambridge university press, 2009.

\bibitem{Bauwens2006}
Luc Bauwens, S{\'e}bastien Laurent, and Jeroen~VK Rombouts.
\newblock Multivariate garch models: a survey.
\newblock {\em Journal of applied econometrics}, 21(1):79--109, 2006.

\bibitem{Clark1973}
Peter~K Clark.
\newblock A subordinated stochastic process model with finite variance for
  speculative prices.
\newblock {\em Econometrica: journal of the Econometric Society}, pages
  135--155, 1973.

\bibitem{Andersen2003}
Torben~G Andersen, Tim Bollerslev, Francis~X Diebold, and Paul Labys.
\newblock Modeling and forecasting realized volatility.
\newblock {\em Econometrica}, 71(2):579--625, 2003.

\bibitem{Von2001}
Hans Von~Storch and Francis~W Zwiers.
\newblock {\em Statistical analysis in climate research}.
\newblock Cambridge university press, 2001.

\bibitem{Franzke2012}
Christian Franzke.
\newblock Nonlinear trends, long-range dependence, and climate noise properties
  of surface temperature.
\newblock {\em Journal of Climate}, 25(12):4172--4183, 2012.

\bibitem{Linkenkaer2001}
Klaus Linkenkaer-Hansen, Vadim~V Nikouline, J~Matias Palva, and Risto~J
  Ilmoniemi.
\newblock Long-range temporal correlations and scaling behavior in human brain
  oscillations.
\newblock {\em Journal of Neuroscience}, 21(4):1370--1377, 2001.

\bibitem{Taylor1994}
Stephen~J Taylor.
\newblock Modeling stochastic volatility: A review and comparative study.
\newblock {\em Mathematical finance}, 4(2):183--204, 1994.

\bibitem{Breidt1998}
F~Jay Breidt, Nuno Crato, and Pedro De~Lima.
\newblock The detection and estimation of long memory in stochastic volatility.
\newblock {\em Journal of econometrics}, 83(1-2):325--348, 1998.

\bibitem{Box2015}
George~EP Box, Gwilym~M Jenkins, Gregory~C Reinsel, and Greta~M Ljung.
\newblock {\em Time series analysis: forecasting and control}, page~33.
\newblock John Wiley \& Sons, 2015.

\bibitem{Singh2016}
Ajay Singh and Dinghai Xu.
\newblock Random matrix application to correlations amongst the volatility of
  assets.
\newblock {\em Quantitative Finance}, 16(1):69--83, 2016.

\bibitem{Schafer2010}
Rudi Sch{\"a}fer, Nils~Fredrik Nilsson, and Thomas Guhr.
\newblock Power mapping with dynamical adjustment for improved portfolio
  optimization.
\newblock {\em Quantitative Finance}, 10(1):107--119, 2010.

\bibitem{Marchenko1967}
Vladimir~A Mar{\v{c}}enko and Leonid~Andreevich Pastur.
\newblock Distribution of eigenvalues for some sets of random matrices.
\newblock {\em Sbornik: Mathematics}, 1(4):457--483, 1967.

\bibitem{Livan2018}
Giacomo Livan, Marcel Novaes, and Pierpaolo Vivo.
\newblock {\em Introduction to Random Matrices: Theory and Practice},
  volume~26.
\newblock Springer, 2018.

\bibitem{Guhr2003}
Thomas Guhr and Bernd K{\"a}lber.
\newblock A new method to estimate the noise in financial correlation matrices.
\newblock {\em Journal of Physics A: Mathematical and General}, 36(12):3009,
  2003.

\bibitem{Livan2011}
Giacomo Livan, Simone Alfarano, and Enrico Scalas.
\newblock Fine structure of spectral properties for random correlation
  matrices: An application to financial markets.
\newblock {\em Physical Review E}, 84(1):016113, 2011.

\bibitem{Wilinski2018}
Mateusz Wilinski, Yuichi Ikeda, and Hideaki Aoyama.
\newblock Complex correlation approach for high frequency financial data.
\newblock {\em Journal of Statistical Mechanics: Theory and Experiment},
  2018(2):023405, 2018.

\bibitem{Biroli2007}
Giulio Biroli, Jean-Philippe Bouchaud, and Marc Potters.
\newblock The student ensemble of correlation matrices: Eigenvalue spectrum and
  kullback-leibler entropy.
\newblock {\em Acta Physica Polonica B}, 38(13), 2007.

\bibitem{Abul2009}
AY~Abul-Magd, Gernot Akemann, and P~Vivo.
\newblock Superstatistical generalizations of wishart--laguerre ensembles of
  random matrices.
\newblock {\em Journal of Physics A: Mathematical and Theoretical},
  42(17):175207, 2009.

\bibitem{Laloux1999}
Laurent Laloux, Pierre Cizeau, Jean-Philippe Bouchaud, and Marc Potters.
\newblock Noise dressing of financial correlation matrices.
\newblock {\em Physical review letters}, 83(7):1467, 1999.

\bibitem{Plerou2002}
Vasiliki Plerou, Parameswaran Gopikrishnan, Bernd Rosenow, Luis A~Nunes Amaral,
  Thomas Guhr, and H~Eugene Stanley.
\newblock Random matrix approach to cross correlations in financial data.
\newblock {\em Physical Review E}, 65(6):066126, 2002.

\bibitem{Mandelbrot1997}
Benoit~B Mandelbrot.
\newblock The variation of certain speculative prices.
\newblock In {\em Fractals and Scaling in Finance}, pages 371--418. Springer,
  1997.

\bibitem{Sharpe1964}
William~F Sharpe.
\newblock Capital asset prices: A theory of market equilibrium under conditions
  of risk.
\newblock {\em The journal of finance}, 19(3):425--442, 1964.

\bibitem{Merton1973}
Robert~C Merton.
\newblock An intertemporal capital asset pricing model.
\newblock {\em Econometrica: Journal of the Econometric Society}, pages
  867--887, 1973.

\bibitem{Borghesi2007}
Christian Borghesi, Matteo Marsili, and Salvatore Miccich{\`e}.
\newblock Emergence of time-horizon invariant correlation structure in
  financial returns by subtraction of the market mode.
\newblock {\em Physical Review E}, 76(2):026104, 2007.

\bibitem{Burda2005}
Zdzis{\l}aw Burda, Jerzy Jurkiewicz, and Bart{\l}omiej Wac{\l}aw.
\newblock Spectral moments of correlated wishart matrices.
\newblock {\em Physical Review E}, 71(2):026111, 2005.

\bibitem{Bloemendal2016}
Alex Bloemendal, Antti Knowles, Horng-Tzer Yau, and Jun Yin.
\newblock On the principal components of sample covariance matrices.
\newblock {\em Probability theory and related fields}, 164(1-2):459--552, 2016.

\bibitem{Diana2002}
Giancarlo Diana and Chiara Tommasi.
\newblock Cross-validation methods in principal component analysis: a
  comparison.
\newblock {\em Statistical Methods and Applications}, 11(1):71--82, 2002.

\bibitem{Theil1958}
Henry Theil.
\newblock {\em Economic forecasts and policy}.
\newblock North-Holland, 1958.

\bibitem{Dietrich1997}
Claude~R Dietrich and Garry~N Newsam.
\newblock Fast and exact simulation of stationary gaussian processes through
  circulant embedding of the covariance matrix.
\newblock {\em SIAM Journal on Scientific Computing}, 18(4):1088--1107, 1997.

\bibitem{Mantegna1999}
Rosario~N Mantegna.
\newblock Hierarchical structure in financial markets.
\newblock {\em The European Physical Journal B-Condensed Matter and Complex
  Systems}, 11(1):193--197, 1999.

\bibitem{Musmeci2015}
Nicolo Musmeci, Tomaso Aste, and Tiziana Di~Matteo.
\newblock Relation between financial market structure and the real economy:
  comparison between clustering methods.
\newblock {\em PloS one}, 10(3):e0116201, 2015.

\bibitem{Musmeci2015b}
Nicolo Musmeci, Tomaso Aste, and Tiziana Di~Matteo.
\newblock Risk diversification: a study of persistence with a filtered
  correlation-network approach.
\newblock {\em Journal of Network Theory in Finance}, 1(1):77--98, 2015.

\bibitem{Musmeci2016}
Nicolo Musmeci, Tomaso Aste, and Tiziana Di~Matteo.
\newblock Interplay between past market correlation structure changes and
  future volatility outbursts.
\newblock {\em Scientific Reports 6}, 6:36320, 2016.

\bibitem{Lancichinetti2008}
Andrea Lancichinetti, Santo Fortunato, and Filippo Radicchi.
\newblock Benchmark graphs for testing community detection algorithms.
\newblock {\em Physical review E}, 78(4):046110, 2008.

\bibitem{Palla2005}
Gergely Palla, Imre Der{\'e}nyi, Ill{\'e}s Farkas, and Tam{\'a}s Vicsek.
\newblock Uncovering the overlapping community structure of complex networks in
  nature and society.
\newblock {\em Nature}, 435(7043):814, 2005.

\bibitem{Micciche2013}
Salvatore Miccich\'{e}.
\newblock Empirical relationship between stocks cross-correlation and stocks
  volatility clustering.
\newblock {\em Journal of Statistical Mechanics: Theory and Experiment},
  2013(05):P05015, 2013.

\bibitem{Bun2016}
Jo{\"e}l Bun, Jean-Philippe Bouchaud, and Marc Potters.
\newblock On the overlaps between eigenvectors of correlated random matrices.
\newblock {\em arXiv preprint arXiv:1603.04364}, 2016.

\bibitem{Karhunen1994}
Juha Karhunen and Jyrki Joutsensalo.
\newblock Representation and separation of signals using nonlinear pca type
  learning.
\newblock {\em Neural networks}, 7(1):113--127, 1994.

\bibitem{Bouchaud2009}
Jean-Philippe Bouchaud and Marc Potters.
\newblock Financial applications of random matrix theory: a short review.
\newblock {\em arXiv preprint arXiv:0910.1205}, 2009.

\bibitem{Markowitz1952}
Harry Markowitz.
\newblock Portfolio selection.
\newblock {\em The journal of finance}, 7(1):77--91, 1952.

\bibitem{ICB}
ftserussel.
\newblock Industry classification benchmark (icb), 2017.

\bibitem{Tibshirani1996}
Robert Tibshirani.
\newblock Regression shrinkage and selection via the lasso.
\newblock {\em Journal of the Royal Statistical Society. Series B
  (Methodological)}, pages 267--288, 1996.

\bibitem{Friedman2010}
Jerome Friedman, Trevor Hastie, and Rob Tibshirani.
\newblock Regularization paths for generalized linear models via coordinate
  descent.
\newblock {\em Journal of statistical software}, 33(1):1, 2010.

\bibitem{Draper2014}
Norman~R Draper and Harry Smith.
\newblock {\em Applied regression analysis}, volume 326.
\newblock John Wiley \& Sons, 2014.

\bibitem{Majumdar2012}
Satya~N Majumdar and Pierpaolo Vivo.
\newblock Number of relevant directions in principal component analysis and
  wishart random matrices.
\newblock {\em Physical review letters}, 108(20):200601, 2012.

\end{thebibliography}

\nocite{*}

\end{document}